%% file: main.tex
\documentclass[a4paper]{IEEEtran}
\usepackage{cite}
\usepackage{amsthm,amsmath,amssymb,amsfonts}
\usepackage{algorithmic}
\usepackage{graphicx}
\usepackage{booktabs, makecell}
\usepackage{threeparttable}
\usepackage{pgfplots}
\usepackage{enumerate}
\usepackage{multirow}
\usepackage[acronym,shortcuts]{glossaries}
\usepackage{hyperref}
\usepackage{tabularx}
\usepackage{url}

\usepackage{tikz}
\definecolor{greycolor}{cmyk}{0,0,0,.8}
\definecolor{accessblue}{cmyk}{0, 1, 1, 1}
\usetikzlibrary{arrows,automata, positioning}
\usetikzlibrary{calc, chains,
                fit,
                positioning,
                shapes}
\usetikzlibrary{patterns}   
\usepackage{color,xcolor,soul}

\definecolor{mybluegray}{rgb}{0.4,0.6,0.8}

\begin{document}

\title{MAGIC: A Method for Assessing Cyber Incidents Occurrence}

\author{\IEEEauthorblockN{Massimo Battaglioni, Giulia Rafaiani, Franco Chiaraluce, Marco Baldi}\\
\IEEEauthorblockA{Polytechnic University of Marche,\\ Department of Information Engineering,\\ Ancona, Italy\\
e-mail: \{m.battaglioni, g.rafaiani, f.chiaraluce, m.baldi\}@univpm.it}
\thanks{The material in this paper was presented in part at the AEIT 2021 International Annual Conference \cite{RafaianiAEIT}. This research was supported in part by the ``Cyber Risk Assessment Models and Algorithms (CybeRAMA)'' project (Ref. 2019.0421), funded by the Cariverona Foundation within the ``Research and Development 2018'' call (\url{https://cyberama.dii.univpm.it/}).
}}

\maketitle

\begin{abstract}
The assessment of cyber risk plays a crucial role for cybersecurity management, and has become a compulsory task for certain types of companies and organizations.
This makes the demand for reliable cyber risk assessment tools continuously increasing, especially concerning quantitative tools based on statistical approaches.
Probabilistic cyber risk assessment methods, however, follow the general paradigm of probabilistic risk assessment, which requires the magnitude and the  likelihood of  incidents as inputs.
Unfortunately, for cyber incidents, the likelihood of occurrence is hard to estimate based on historical and publicly available data; so, expert evaluations are commonly used, which however leave space to subjectivity.
In this paper, we propose a novel  probabilistic model, called MAGIC (Method for AssessinG cyber Incidents oCcurrence), to compute the likelihood of occurrence of a cyber incident, based on the evaluation of the cyber posture of the target organization.
This allows deriving tailor-made inputs for probabilistic risk assessment methods, like HTMA (How To Measure Anything in cybersecurity risk), FAIR (Factor Analysis of Information Risk) and others, thus considerably reducing the margin of subjectivity in the assessment of cyber risk. We corroborate our approach through a qualitative and a quantitative comparison with several classical methods. 
\end{abstract}

\section{Introduction}\label{sec:intro}

The massive exploitation of data and information systems in companies and organizations is motivating an increasing attention to cybersecurity and its management.
One of the main pillars upon which cybersecurity management relies is cyber risk assessment, for which a plethora of standards and models exist. Risk does not have a unique definition, but according to the NIST  (National Institute of Standards and Technology) \cite{sp800-30}, it is \emph{a measure of the extent to which an entity is threatened by a potential circumstance or event, and is typically a function of: (i) the adverse impacts that would arise if the circumstance or event occurs; and (ii) the likelihood of occurrence}.
A unique method for cyber risk assessment does not exist, but international standards provide general guidelines which should be followed when designing a cyber risk assessment method. For example, the ISO  (International Standard Organization) unpacks risk assessment into risk identification, analysis and evaluation \cite{31000,27005}. In the risk identification phase, the critical services are identified and the threats and vulnerabilities they could cause are determined. Risk analysis, instead, is needed to determine the likelihood of occurrence and the impact of these threats \cite{isra,27005}. Finally, in the risk evaluation phase the obtained results are compared with some pre-established risk acceptance criteria \cite{31000,27005}. 
In this paper we focus on the second one of the aforementioned phases, that is, risk analysis and likelihood estimation.

Existing cyber risk assessment approaches can be grouped into quantitative and qualitative methods. In the former ones, the risk analysis phase is carried out through numerical evaluations, mostly based on probability theory, as in classical PRA (probabilistic risk assessment) \cite{Stamatelatos2000}.
Widely speaking, these approaches have some potential advantages, like their robustness, reproducibility, and comparability of results. While it seems generally reasonable to rely on past events for risk analysis, in the specific case of cyber incidents these data may be unavailable or incomplete and, thus, probabilistic methods must resort to the help of experts, which makes them barely practical in real-case scenarios \cite{sp800-30}. 

Qualitative methods, in turn, exploit a nonnumerical approach and, therefore, they are often simpler to implement and interpret. However, their results are scarcely reproducible and comparable and, above all, they have an intrinsic nature of subjectivity.

\subsection{Contribution}

To the best of our knowledge, a probabilistic method enabling the computation of the likelihood of occurrence of cyber incidents, based on the posture of the organization rather than on subjective expert estimates, is currently missing.
To fill this gap, we propose a method called MAGIC (Method for AssessinG cyber Incidents oCcurrence), which is a probabilistic model to quantitatively estimate the likelihood of a cyber incident for a specific organization, starting from qualitative assessment approaches based on questionnaires. 
Owing to its own nature, MAGIC should be seen as a tool to be used in conjunction with existing probabilistic cyber risk assessment methods, to make their results more reproducible and less subjective.

For this purpose, we consider four indicators representing the \emph{awareness} of the employees, the \emph{maturity}, the \emph{complexity}, and the \emph{attractiveness} of the target organization. 
These aspects are commonly qualitatively assessed through questionnaires, so we estimate the values of the corresponding indicators starting from questionnaires concerning the target organization, provided by one or more assessors.
Then, the four indexes are combined, using a probabilistic quantitative approach, in order to find the likelihood of occurrence of a given type of cyber incident. In particular, we consider two scenarios: in the first simplified scenario, we obtain as output the estimated frequency of occurrence of cyber incidents in a given time period (and the associated likelihood); in the second scenario, we obtain the probability that the organization will face exactly one successful cyber incident in a given time period. Indeed, in the latter case we assume that, after one cyber incident, the organization will change its posture and thus a new assessment needs to be performed. 

MAGIC provides inputs that can be used in conjunction with classical PRA methods such as HTMA (How To Measure Anything in cybersecurity risk) \cite{htma}, FAIR (Factor Analysis of Information Risk) \cite{fair}, as well as many others. In the former method, it is assumed that the frequency of adverse events follows a log-normal distribution, whose mean and variance is not directly related to the organization posture. 
Similarly, in the FAIR method, for a given adverse event, an expert is needed to estimate the minimum, maximum and most likely values for its frequency of occurrence. 
In the case of cyber events, these processes often become subjective, due to the lack of reliable historical data concerning each specific type of cyber incident.
Our target is to reduce such a subjectivity as much as possible, by following and extending the approach in \cite{santini2019,RafaianiAEIT}.

In short, according to the proposed method, the main effort required from the target organization is to provide information regarding the state of their technological systems and protection measures: expert evaluations are no longer required and the relations among all the components of the considered infrastructure do not need to be figured out. 
We consider questionnaires based on international cybersecurity standards and frameworks, which are comprehensive and widely recognized, but MAGIC is general and can be applied with other types of questionnaires.

A basic version of our model has been introduced in \cite{RafaianiAEIT}. Even though the approach proposed in this paper has some similarities with that in \cite{RafaianiAEIT}, since the final goal of both of them is to estimate the likelihood of occurrence of cyber incidents based on the organization posture, there are some profound differences between them. The most important ones are described next:
\begin{itemize}
    \item the approach in \cite{RafaianiAEIT} does not take the awareness of the employees into account as a determinant factor for risk assessment, which is instead considered in the model we propose;
    \item we provide the set of controls to be used for complexity assessment;
    \item the approach in \cite{RafaianiAEIT} stands as an independent method for probabilistic assessment, whereas MAGIC is a transversal tool to provide tailor-made inputs for existing probabilistic cyber risk assessment methods; owing to this, we are also able to report a larger number of numerical results;
    \item in this paper, the final likelihood is computed differently from that in \cite{RafaianiAEIT}, where the probability that the organization faces a certain number of cyber attacks is not considered in the computation. Through the novel approach, instead, the likelihood of a cyber incident can be easily converted into a frequency, comparable to that obtained through expert evaluations in existing methods;
    \item differently from \cite{RafaianiAEIT}, in this paper we also consider the practical scenario in which the target organization does not immediately realize that a cyber incident has occurred.
\end{itemize}

\subsection{Paper organization}

The paper is organized as follows. In Section \ref{sec:relatedworks} we provide a high-level description of many related works. In Section \ref{sec:quantimet} we recall the basic functioning of some probabilistic cyber risk assessment methods. In Section \ref{sec:modelcompo} the components of the cyber incident occurrence model we propose are discussed. In Section \ref{sec:merging} we show how our approach can be combined with known probabilistic methods in order to overcome their main limitations when dealing with cyber risks. In Section \ref{sec:Results} we present some numerical results, aimed at validating  the effectiveness of the proposed approach. In Section \ref{sec:comdiscuss} we provide both a quantitative and a qualitative comparison of our method with other approaches. Finally, Section \ref{sec:concl} concludes the paper.

\section{Related works} \label{sec:relatedworks}

A plethora of approaches for cyber risk assessment can be found in existing literature. Besides those proposed by national and international organizations (like the mentioned ISO/IEC (International Standard Organization/International Electrotechnical Commission) 27005:2018 \cite{27005} and NIST SP (Special Publication) 800-30 \cite{sp800-30}), others have been introduced by public and private organizations, like EBIOS (Expression des Besoins et Identification des Objectifs de S\'ecurit\'e) \cite{ebios}, CRAMM (Central Computer and Telecommunications Agency Risk Analysis and Management Method) \cite{cramm}, OCTAVE (Operationally Critical Threat, Asset, and Vulnerability Evaluation) \cite{octave}, MEHARI (MEthod for Harmonized Analysis of RIsk) \cite{mehari}, MAGERIT (Risk Analysis and Management Methodology for Information Systems) \cite{magerit}, IRAM2 (Information Risk Assessment version 2) \cite{iram2}, IT-Grundschutz  \cite{grund}, and CORAS  \cite{coras}. An extensive and critical literature review can be found in \cite{survey,meta-survey}. A lot of attention has been devoted to solving the problem of estimating the likelihood of occurrence of a threat and the corresponding impact. For example, several methods have been proposed using different techniques like Bayesian networks \cite{bayesian}, attack path graphs \cite{lisra}, fuzzy logic \cite{fuzzy}, probabilistic model checking \cite{prm}, vulnerability assessments \cite{cvss_attackgraphs}, Monte Carlo simulations \cite{htma,fair}, and others. 

Next we provide a brief description of some of the aforementioned methods, highlighting the differences with MAGIC as well as the possible common aspects. 
A deeper comparison is then carried out in Section \ref{sec:comdiscuss}, after the description of our method.

\begin{itemize}
\item \textbf{ISO/IEC 27005:2018} \cite{27005}: is an international framework for managing information risks. It describes all processes for risk management, including risk assessment. In this case, the estimation of the likelihood can be performed in a qualitative, quantitative, or hybrid way. However, the ISO/IEC 27005 standard only provides guidelines for doing it, without describing any specific practical method. 
\item \textbf{NIST SP 800-30} \cite{sp800-30}: is a guide for conducting risk assessment. In order to determine the likelihood of occurrence of a security incident, this method identifies all the potential vulnerabilities and the probability of their exploitation. The likelihood is then described using a qualitative or semi-quantitative scale. Such an approach is opposed to the numerical one we propose, since it leaves space to subjectivity.
\item \textbf{EBIOS} \cite{ebios}: is a scenario-based approach for risk management that relies on the establishment of a strong link among different stakeholders. It uses a modular approach for identifying risk causes. However, all the phases considered in this method, including risk assessment, are the result of security debates among the team and, therefore, they are subjective.
\item \textbf{CRAMM} \cite{cramm}: is a qualitative method for risk assessment. It measures risks as the product of asset, threat, and vulnerability values. It uses trained experts. Threats and vulnerabilities are not exhaustively assessed, but the assessor can choose among different predefined threat/asset and threat/impact combinations. Relying on structured questionnaires, and/or on the expertise of the assessor, the method determines the likelihood of threats and vulnerabilities; however, differently from MAGIC, it does not directly link those likelihoods with the likelihood of occurrence of a cyber incident.
\item \textbf{OCTAVE} \cite{octave}: is an asset-driven method for assessing information security risks. While the original approach is designed for large organizations, the OCTAVE-S \cite{octaves} and the more recent OCTAVE-Allegro \cite{octaveallegro} versions can be also applied to small and medium enterprises. This method firstly identifies all the assets, and then focuses on the critical ones. For each of them, it determines the related threats, and qualitatively labels their likelihood of occurrence as ``Low'', ``Medium'' or ``High''. 
\item \textbf{MEHARI} \cite{mehari}: is a risk management model with the aim of helping the implementation of ISO/IEC 27005. It performs the risk assessment phase through an audit. This phase includes the identifications of assets, threats and vulnerabilities. The likelihood of occurrence of a threat is qualitatively described using a four levels scale.
\item \textbf{MAGERIT} \cite{magerit}: is an asset-oriented risk management model. Using this model, security professionals evaluate the assets and all the possible threats. Then, they describe the likelihoods of occurrence of those threats using a numerical scale with a limited number of levels. The likelihood is usually evaluated relying on the  annual rate of occurrence of any specific threat for each specific asset, which is substantially estimated through historical data.
\item \textbf{IRAM2} \cite{iram2}: is a qualitative threat-driven method for information risk assessment and treatment. The likelihood of success of a threat is estimated using lookup tables having as input two numerical values describing the strength of the threat and that of the security controls implemented by the organization. Then, using this method, the residual likelihood is evaluated (also, through a lookup table) as a combination of likelihood of success and likelihood that an attacker will try to cause an incident based on the considered threat. 
\item \textbf{IT-Grundschutz} \cite{grund}: is a qualitative method for identifying and assessing security incidents. Using this method, some qualified staff has to identify all possible threats. Then, for each of them this method evaluates the frequency of occurrence using a qualitative scale. This frequency is finally combined with the impact through a risk matrix.
\item \textbf{CORAS} \cite{coras}: is a method for security risk analysis. All the risk assessment phases are performed through structured brainstorming, where people with different backgrounds and competences collaborate to identify assets, vulnerabilities, threats, and the related likelihoods of occurrence. The likelihood assessment can be done both in a qualitative or a quantitative way; however, the evaluation is strictly related to the subjectivity of the assessors.
\item \textbf{LiSRA} \cite{lisra}: is a risk assessment method taking a bidimensional input; on one dimension domain-specific information is required from an expert, while the other dimension is filled by the user, according to the practices of the considered organization. Then, the risk is conventionally obtained as the combination of the probability to have successful attacks and their impact. The probability of success is computed based on attack trees but, differently from our method, the latter need to be entirely computed by external experts. Therefore, MAGIC might be seen, with some adaptations, as an alternative method allowing to bypass the need of experts, rather than a completely different approach.
\item \textbf{Risk analysis based on fuzzy decision theory} \cite{fuzzy}: the first step of this approach is to identify an expert; then, a  taxonomy of events and scenarios has to be defined (second step). Finally, the expert builds a matrix with potential accidents on the rows and possible scenarios on the columns: each entry of the matrix has to be filled with a probability that the accident takes place in a certain scenario. Also in this case, differently from our method, external expert estimates are used. Starting from this matrix, the fuzzy decision theory is then applied, which returns the expected value of the considered option. Also in this case, MAGIC is not in contrast with this method, but can be seen as a variation of it. In fact, in MAGIC the a-priori probabilities can be numerically derived from the posture of the organization, rather than estimated by an expert.
\item \textbf{CVSS-based risk assesment}: the CVSS (Common Vulnerability
Scoring System) \cite{cvss} gives scores to threats exploiting vulnerabilities on the basis of three categories. One can combine this approach with attack graphs \cite{cvss_attackgraphs} in order to derive the vulnerabilities starting from known threat sources, or to bayesian decision networks \cite{bayesian}, in which the attacks are modeled starting from correlated alerts. The probabilities that the considered threats exploit certain vulnerabilities are computed from the scores associated to each of them, according to CVSS. In particular, CVSS 2.0 provides for the use of three types of metrics: base, temporal and environmental. Base metrics represent the intrinsic characteristics of threats exploiting vulnerabilities that are constant over time and user environments; temporal metrics represent the features of threats exploiting vulnerabilities that change over time but not over user environments; environmental metrics are based on the characteristics of threats exploiting vulnerabilities that are unique to a particular user's environment. These methods may be directly comparable to MAGIC, in that the probabilities derive from numerical assessments, even though a subjective component is still required. In particular, in the basic version of CVSS-based risk assessment methods, the probability of occurrence of the $i$-th item in the Common Vulnerabilities and Exposures (CVE) list can be computed as a product of some CVSS metrics, i.e.,
\[
L_i=AV \times AC \times Au \times E \times RC,
\]
where $AV$ is the access vector metric, $AC$ is the access complexity metric, $Au$ is the authentication metric, $E$ is the exploitability metric and $RC$ is the report confidence metric. The exact numerical values of all these metrics need to be chosen in a range of prefixed values, which makes this numerical approach also subjective. 
\end{itemize}

There are several reasons that make many of the aforementioned approaches difficult to apply in real case scenarios. In fact, as explained above, they make risk assessment result in a long process requiring the availability of a significant amount of data. Moreover, the application of these methods often requires the help of an expert assessor external to the organization and able to provide quantitative measures of the cyber risk, which makes them cumbersome and exposed to subjectivity.
MAGIC, instead, allows computing quantitative parameters starting from simple questionnaires, which makes the risk assessment process straightforward and less exposed to subjectivity.

\section{Probabilistic cyber risk assessment: preliminaries}\label{sec:quantimet}

In this section we describe two state-of-the-art probabilistic cyber risk assessment methods based on Monte Carlo simulations, i.e., HTMA \cite{htma} and FAIR \cite{fair}.

\subsection{Probabilistic risk assessment using HTMA}\label{subsec:accatma}

In a nutshell, the HTMA method for cyber risk assessment \cite{htma} is based on the following steps:
\begin{itemize}
    \item definition of the potential cyber threats;
    \item estimation of the likelihood of occurrence and impact of each event;
    \item Monte Carlo simulation for generating the scenarios;
    \item results interpretation.
\end{itemize}

While the likelihood of occurrence of each threat is given by a single value, the impact is associated to a $90\%$ confidence interval, identified by a lower and an upper bound. All these three numerical values are to be determined by external experts.

The Monte Carlo simulation, in each scenario, works as follows:
\begin{enumerate}[1.]
    \item for any threat $i$, a real number $r$  is generated by sampling uniformly at random the range between $0$ and $1$, boundaries included, denoted as $[0,1]$. If $r<L_i$, where $L_i$ is the likelihood of the $i$-th threat, it is assumed that the event has not happened, and vice versa;
    \item the impact of events which did not occur is $0$, whereas the impact of occurred events is obtained by randomly sampling a log-normal distribution of the impacts, obtained according to the values of the given $90\%$ confidence interval;
    \item all the impacts of occurred events are summed, in order to obtain an estimate of the total annual risk.
\end{enumerate}

The results obtained with the Monte Carlo simulation are used to construct the LEC (Loss Exceedance Curve), which corresponds to the graphical representation of the complementary cumulative distribution function of the annualized loss expectancy.

\subsection{Probabilistic risk assessment using FAIR}

The FAIR methodology \cite{fair} can be described through the following four steps:
\begin{itemize}
    \item definition of the scenario under exam and its decomposition into sub-scenarios;
    \item estimation of the parameters for any sub-scenario;
    \item generation of the frameworks through Monte Carlo simulations;
    \item results interpretation.
\end{itemize}

\begin{figure*}[tb]
    \begin{centering}
    \input{Figures/fair}
    \caption{Ontology of the FAIR risk.}
    \label{fig:schemeFAIR}
    \end{centering}
\end{figure*}
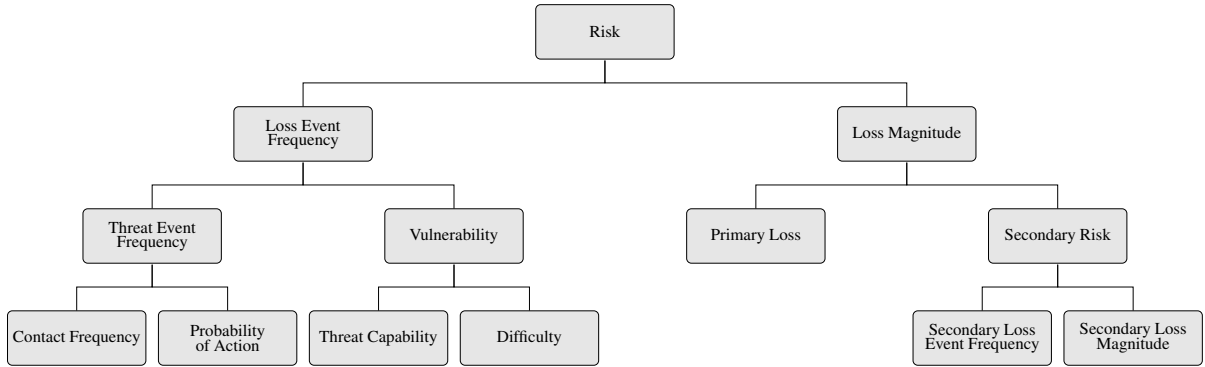

More generally, FAIR defines an ontology for the risk, which is summarized in Fig. \ref{fig:schemeFAIR}. The ontology describes how the assessment of risk can be obtained; in particular, it considers all the risk factors that contribute to the evaluation of the risk and all the relationships between them. Risk factors can be measured and estimated; then, it is possible to calculate risk using mathematical expressions of the relationships among factors.
In essence, the risk is computed as a combination of  LEF (Loss Event Frequency) and Loss Magnitude. In order to facilitate the process of risk evaluation, these two factors can be individually decomposed in other factors that, in their turn, could be further decomposed, as well. This way, the user can assess the risk considering a specific layer of the ontology, according to which factors he is able to estimate. 
The LEF depends on several factors; the most relevant ones for our model  are redefined next: the TEF (Threat Event Frequency) is defined as the frequency with which, in a given time period, the attacker tries to breach the organization; the Vulnerability is the probability of success of any of these breaches. The Loss Magnitude, instead, is the sum of the losses caused by the a certain primary threat event  and of the losses caused by its side-effects (such as, for example, the reaction of secondary stakeholders), which are included in the concept of secondary risk. 

Conventionally, the outputs of the FAIR approach are scatterplots with LEF and Loss Magnitude on the $x$-axis and $y$-axis, respectively, and/or tables summarizing various results of the Monte Carlo simulation.

\section{Cyber incident model}\label{sec:modelcompo}

In this section we introduce the quantitative approach we propose for estimating the likelihood of occurrence of a cyber incident, called MAGIC. First of all, we provide some basic definitions:
\begin{itemize}
\item \textbf{Cyber threat} \cite{sp800-53}: any circumstance or event with the potential to adversely impact organizational operations (including mission, functions, image, or reputation), organizational assets, or individuals through an information system via unauthorized access, destruction, disclosure, modification of information, and/or denial of service. We will simply refer to cyber threats as ``threats'' in the following.
\item \textbf{Cyber attack}: any realized attempt to partially or totally disclose, expose and/or compromise data by a malicious entity. We will simply refer to cyber attacks as ``attacks'' in the following.
\item \textbf{Cyber incident}: any cyber attack which had success. 
\end{itemize}

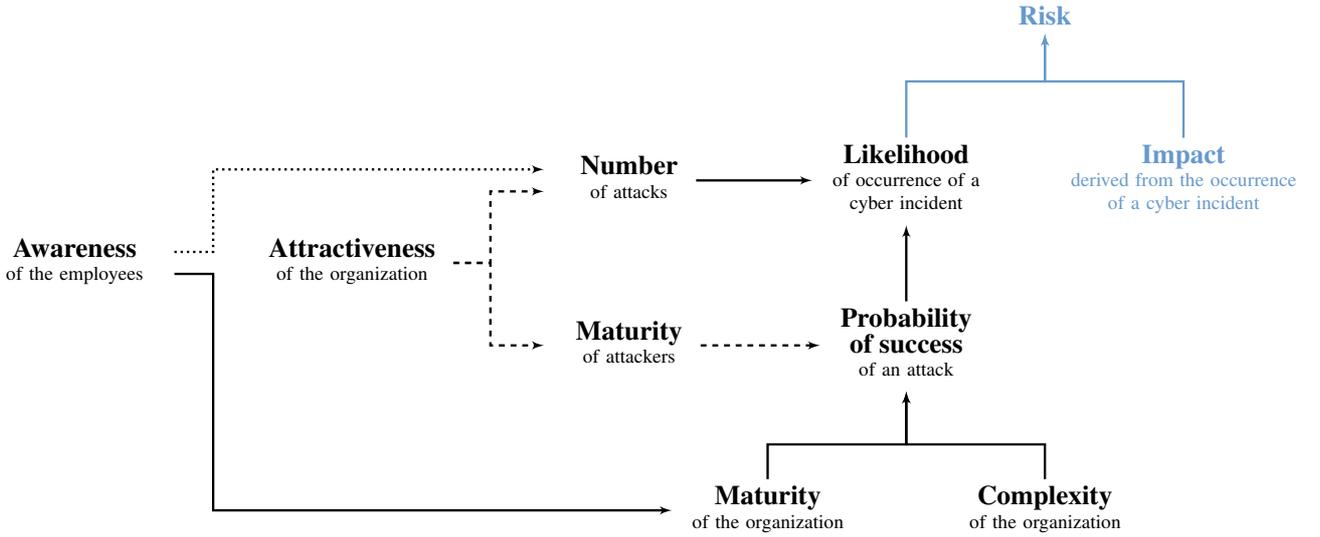
\begin{figure*}[tb]
    \begin{centering}
    \input{Figures/model}
    \caption{Relations among the components of MAGIC.}
    \label{fig:scheme}
    \end{centering}
\end{figure*}

The proposed model, with all the key parameters, which we will discuss next, is schematized in Fig. \ref{fig:scheme}. We need to distinguish between two types of threats: those coming from external threat agents and those happening as a consequence of human misbehavior, \textit{i.e.}, non-malicious threats. In Fig. \ref{fig:scheme}, the dashed arrows are referred to the former scenario, the dotted arrows describe the latter, whereas the solid arrows are valid for both scenarios.  
The model components in black are described in detail in the following subsections, while those in blue are not among the objects of our theoretical analysis, but they are taken into account in Section \ref{sec:Results}, where numerical results are provided. 

\subsection{Awareness of the employees}
\label{subsec:awareness}
The awareness of the employees can be defined as their level of consciousness about the cybersecurity risks. It can be directly related to the training programs the organization supplies to its workers. 

In order to practically evaluate the awareness, it is possible to consider a list of best practices devoted to this issue, as that in the NIST SP 800-53 \cite{sp800-53}, or to rely on specific subsets of controls proposed in cybersecurity frameworks; for example, Control 14 in the CIS (Center for Internet Security) Controls \cite{cis} deals with security awareness and training programs. The evaluation of the awareness parameter can be carried out in different ways. For example, the assessor  can determine if every identified control is fully implemented by the organization or not, simply associating to it a \textit{Yes} or \textit{No} answer. Another approach can be using a scale in order to determine at which degree each considered control is implemented and, therefore, assigning better ratings as the level of implementation completeness of the control increases. With both these approaches, an \textit{N/A} option needs to be included; this should be used when one or more controls are considered to be not applicable in the context under examination. Then, according to the results of the assessment, a numerical score should be assigned to every control. For example, using the first of the approaches stated above, a score equal to $1$ may be assigned to all the controls with \textit{Yes} as an answer, while $0$ may be assigned to the controls answered with \textit{No}. Instead, when using the second approach, one may consider as many scores as the number of the possible implementation levels. In other words, if the scale used for the implementation evaluation has $5$ possible levels, the score that can be assigned to each control can be, for example, a number between $0$ and $4$. But, obviously, other choices are possible. In the following, we denote by $s_{\max}$ the maximum score value (so, $s_{\max} = 1$ or $s_{\max} = 4$ in the mentioned examples). Moreover, a weight may be assigned to every control. The weights are given mainly according to the aspects that the organization wants to stress out, so they are not mandatory. In fact, an organization may want to focus its analysis on some specific training strategies, while it may not be interested in other ones; in this case, it can assign higher (or lower) weights to the controls that it considers more crucial (or less relevant). Finally, a weighted average is calculated in order to obtain a final value for the awareness. Note that the \textit{N/A} controls should not be counted in the average.

In this paper, we consider the awareness index as a number between $0$ and $10$. Let $E$ be the total number of considered controls, $s_i \in [0, s_{\max}]$ and $a_i \geq 0$ the score and the weight associated to the $i$-th control,  respectively; the awareness index is computed as
\begin{equation}
    \mathrm{Awareness \hspace{2pt} Index} = \mathrm{AI} =  \frac{\sum_{i=1}^{E}s_i \times a_i}{\sum_{i=1}^{E}a_i} \times \frac{10}{s_{\max}}.
\end{equation}

Notice that, as shown in Fig. \ref{fig:scheme}, the awareness directly influences the maturity of the organization, formally defined in Section \ref{subsec:maturity}. This holds for both malicious and non-malicious threats.  Moreover, when specifically dealing with non-malicious threats, the awareness is inversely related to the number of threats. In other words, if employees are well-trained about cyber risks, we can assume that the organization will suffer fewer potential threats, since the employees are less prone to causing cyber incidents.

\subsection{Maturity of the organization} \label{subsec:maturity}

The maturity of an organization can be defined as the level of implementation of all practices and procedures that the organization executes in order to reduce the risk of receiving cyber attacks which may cause security breaches, data leakage, denial of service, and so on. As mentioned in Section \ref{subsec:awareness}, awareness concurs to determine maturity (being one of its most relevant components) but other issues need to be taken into account, too. 
The evaluation of this further part  of the maturity can be done by assessing the organization compliance to the security controls proposed by one or more cybersecurity frameworks. The frameworks chosen as the reference ones will influence the area of application and, therefore, the kind of risk the organization is going to evaluate. An organization, in fact, may be interested in assessing the risk of suffering breaches, or the risk of losing, as a consequence of an attack, data confidentiality and/or integrity and/or availability, or may be interested in assessing the risk related to a combination of them both. The security frameworks that can be used as reference are, for example, the one proposed by CIS in \cite{cis} for assessing cybersecurity risk, or the set of controls introduced by the ENISA  (European Union Agency for Cybersecurity) in \cite{enisa} for assessing data protection risk, or the NIST Cybersecurity Framework \cite{csf} for assessing both cybersecurity and data protection risks. Therefore, first of all, the organization should choose the kind of risk it intends to assess and, according to the choice, select one or multiple appropriate security frameworks as reference. The next step is evaluating if and at which level all the security controls listed in the reference framework are implemented within the organization. Using the controls of cybersecurity frameworks to assess the maturity will result in an accurate picture of the actual cyber posture of the organization. In fact, since the security frameworks are usually considered as a set of best practices, their use will lead to a robust and complete mapping of the area of interest. Moreover, since the security frameworks are continuously updated, the maturity can be dynamically assessed, simply repeating the evaluation as soon as new versions of the frameworks are released.

Because of the need to combine the mentioned different aspects, the maturity index evaluation is performed through a two-step procedure. The first step is equivalent to that described in the previous section for the evaluation of the awareness index. Clearly, the controls of the chosen framework about cybersecurity awareness and training programs must be excluded from the list used for this part of the evaluation. Thus, after answering to all the remaining controls, a score is associated to any answer. The \textit{N/A} option should  be included as well. Also in this case, it is possible to assign a weight to each control, according to its relevance for the kind of assessment the organization wants to perform. Let $T$ be the total number of considered controls, $s'_i \in [0, s'_{\max}]$ and $a'_i \geq 0$ respectively the score and the weight associated to the $i$-th control; a weighted average is computed as follows
\begin{equation}
    \mathrm{M} = \frac{\sum_{i=1}^{T}s'_i \times a'_i}{\sum_{i=1}^{T}a'_i} \times \frac{10}{s'_{\max}}.
\end{equation}

Then, in the second step of the procedure, the maturity index is eventually obtained as a weighted average of the awareness index and $\mathrm{M}$. In this case, $\mathrm{AI}$ and $\mathrm{M}$ are weighted by $E/(E+T)$ and $T/(E+T)$, respectively. We finally obtain

\begin{equation}
    \mathrm{Maturity \hspace{2pt} Index} = \frac{\mathrm{AI} \times E + \mathrm{M} \times T}{E+T}.
\end{equation}
We observe that, according to this definition, the maturity index is a real number ranging between $0$ and $10$.

Note that the maturity of the organization will directly influence the probability of success of an attack. In fact, a higher maturity means that the organization has addressed more attention to cybersecurity practices and, therefore, the probability for an attack attempt to be successful will decrease, and vice versa.

\subsection{Complexity of the organization}

The complexity of an organization can be defined as the measurement of the intricacy of its technological infrastructure and of how the processes, the activities, and the services are managed. The concept that the risk does not only depend on the maturity of the organization, but also on its complexity is introduced in \cite{cis}, where three IGs  (Implementation Groups) are defined. Following that approach, every organization should identify itself, mainly according to its dimension, in one of the proposed IGs and, therefore, should implement a specific subset of security controls. This is due to the fact that smaller organizations are usually expected to be less exposed to threats, when compared to larger organizations. However, the dimension should not be considered as the only discriminating parameter. For this reason, starting from the controls proposed in \cite{acat} for the evaluation of the inherent risk, we have identified a set of punctual and specific controls useful to assess the inherent complexity of an organization\footnote{The set of controls considered in this paper for the complexity assessment can be found at \url{https://github.com/secomms/cyber-risk-assessment}.}. Coherent with the above considerations, the controls we address do not consider only the dimension of the organization, but they try to identify all possible critical points of  hardware, software, networks, and facilities. More precisely, we have grouped all the considered controls into five categories: Networks and Infrastructure, IP (Internet Protocol) network technologies, Applications, Services and IT (Information Technology) department. Therefore, beside the number of employees, the assessment includes the number and the characteristics of the components (physical and software systems) and their interconnections, the number of services and their characteristics, and the entropy of the IT system management. In order to facilitate the assessment, for each control we have identified five possible guided answers. The answers are associated to \textit{Very Low}, \textit{Low}, \textit{Medium}, \textit{High}, and \textit{Very High} complexity.

In order to practically assess the complexity, the process is equivalent to the maturity assessment. After evaluating all the controls, a score is associated to every evaluation. The \textit{N/A} option should  be included as well. Also in this case, it is possible to assign a weight to each control, according to its relevance for the kind of assessment the organization wants to perform. Then, a weighted average is calculated in order to obtain a complexity index for each one of the five categories considered above. Similarly to the previous indexes, we consider the complexity index as a number between $0$ and $10$. Let $C_j$ be the number of controls included in the category $j$, $s''_{i,j} \in [0, s''_{j,\max}]$ and $a''_{i,j} \geq 0$ respectively the score and the weight associated to the $i$-th control of the $j$-th category; the complexity index of the category $j$ is computed as
\begin{equation}
    \mathrm{CI}(j) = \frac{\sum_{i=1}^{C_j}s''_{i,j} \times a''_{i,j}}{\sum_{i=1}^{C_j}a''_{i,j}} \times \frac{10}{s''_{j,\max}}.
\end{equation}
Finally, a global \textit{complexity index} for the organization is computed as the weighted average of the five complexity indexes computed before; in this case, the weight associated to each category is simply computed as the number of controls included in that category divided by the total number of controls, i.e., $b_j = C_j/(\sum_{j=1}^{5}C_j)$. Then, we have
\begin{equation}
    \mathrm{Complexity \hspace{2pt} Index} = \sum_{j=1}^{5}(\mathrm{CI}(j) \times b_j).
\end{equation}

We observe that the same approach with the two weighted averages could be used for the maturity assessment when the security controls chosen for the evaluation are divided into categories.

Note that the complexity of the organization will inversely influence the probability of success of an attack. In fact, a higher complexity means that the technological infrastructure of the organization is more intricate and, therefore, it is more prone to being successfully targeted. In other words, the probability for an attack attempt to be successful will increase for increasing complexities, and vice versa, according to the law that will be described in Section \ref{subsec:prothr}.

\subsection{Attractiveness of the organization}\label{subsec:attractiveness}

The attractiveness of an organization can be defined as the level of interest the organization causes in potential attackers. The attractiveness depends on several factors, as the type of business, the kind and the amount of data the organization manages, etc. The insertion of this parameter in our model is due to the assumption that cyber criminals will likely attack organizations from which they can obtain larger profits. The dimension of the organization does not affect its attractiveness. In fact, for example, small organizations may operate in critical environments and/or may process a large amount of sensitive data, making them more attractive to cyber criminals if compared to larger organizations with a limited technological value.

In order to practically estimate the attractiveness, we have examined the data proposed in cybersecurity reports like \cite{clusit}. In particular, we have considered the number of attacks any type of organization has suffered in one year, with respect to the total number of attacks analyzed in the report. According to the given data, we have classified the type of organization as reported in Table \ref{table:attract}.

\begin{table}[ht!]
\centering
   \caption{Classification of the organizations as a function of the percentage $\pi$ of attacks received with respect to the total number.}
\begin{tabular}{cc}
        \hline
         Percentage of attacks & Type of organization \\
         \hline\hline
         $\pi < 1.25\%$ & \textit{Very lowly} attractive \\
         \hline
         $1.25\% \leq \pi < 2.5\%$ & \textit{Lowly} attractive \\
         \hline
         $2.5\% \leq \pi < 5\%$ & \textit{Averagely} attractive \\
         \hline
         $5\% \leq \pi < 10\%$ & \textit{Highly} attractive \\
         \hline
         $\pi \geq 10\%$ & \textit{Very highly} attractive \\
         \hline
        
        \end{tabular}
    \label{table:attract}
\end{table}

Therefore, during the assessment, the organization should simply identify itself in one of the proposed types of business, and the attractiveness will be assigned consequently.

Note that the attractiveness will influence both the number of attacks and the maturity of attackers. In fact, we assume that a highly attractive organization will likely suffer more and more structured attacks if compared to a lowly attractive organization.

\subsection{Maturity of attackers}

Given the definition of cyber attack proposed in Section \ref{sec:modelcompo}, let us devise the attacker model. As mentioned above, we should  distinguish between attacks coming from malicious attackers, and threats deriving from the lack of awareness of the employees. For the probabilistic model we devise, however, there is no need to mathematically distinguish between these cases. Thus, for the sake of simplicity, in the rest of the paper we will also refer to non-malicious threats as ``attacks'', keeping in mind that they come from non-malicious employees, who assume the role of unaware attackers in the model.

In general, the organization might be targeted by multiple attackers, but we assume that they cannot conduct more than one attack in the same time slot $\Delta t$. The value $\Delta t$ can be chosen sufficiently small, so that attacks performed in close periods of time can be distinguished. Moreover, under this assumption, the identity of the attacker does not play a relevant role and, therefore, we will generically refer to ``the attacker''. Another assumption that we make, in order to keep the analysis feasible, is that different attack attempts are not correlated. In other words, each attack attempt does not depend on previous attack attempts, and its outcome does not influence future attack attempts. Clearly, this hypothesis may not always be verified and, in the scenario where the assumptions are too optimistic, our model provides a lower bound to the likelihood of the adverse events, rather than an estimate.

As shown in Fig. \ref{fig:scheme}, the attractiveness is related to the maturity of the attackers: attractive organizations will more likely face more structured attacks, and vice versa. The maturity of the attackers, in its turn, influences the probability of success of an attack, described in the next section.

\subsection{Probability of success of an attack}\label{subsec:prothr}

Obviously (and luckily) attacks are not always successful and, therefore, more than one attempt may be needed before breaking through the organization defenses. The single attack attempt is thus associated to a probability of success. In particular, the computation of the probability of success of a single attack takes into account the four key indexes described above. Considering the maturity of the organization as a variable $x$, we need a function that decreases when $x$ increases, since we expect more mature organizations to be more resistant to cyber attacks, and vice versa. Furthermore, we do not expect this trend to be linear, since slight improvements of the maturity of a very immature organization may not be sufficient to significantly decrease the probability of success of an attack and, similarly, very mature organizations should not need to further improve significantly their posture, since the probability of an attacker breaching them may already be low enough. A function that may fit well this scenario is the logistic function \cite{logistic}
\begin{equation}
    f(x)=\frac{K}{1+e^{-B(x-t)}},
    \label{eq:log}
\end{equation}
where $K$ is the \textit{saturation level}, \textit{i.e.}, the upper horizontal asymptote that limits the curve's maximum value; $t$ is the \textit{midpoint}, \textit{i.e.}, the value of $x$ corresponding to half the saturation level and  $B$ is the \textit{growth rate}, \textit{i.e.}, the steepness of the curve. In order to obtain the aforementioned trend, $B$ must be chosen negative. According to \eqref{eq:log}, $f(x)$ has a lower asymptote equal to $0$.

It is possible to extend \eqref{eq:log}, by considering a lower asymptote different from $0$ and a non-symmetric shape, thus obtaining the so-called \textit{generalized logistic function}  \cite{genlog}
\begin{equation}
    f(x)=A+\frac{K-A}{(1+Q \times e^{-B(x-x_0)})^{1/\nu}},
    \label{eq:genlog}
\end{equation}
where $A$ is now the \textit{lower asymptote}, $Q$ is a variable, related to $f(0)$, that influences the inflection point,  $f(x_0)=A+\frac{K-A}{(1+Q)^{1/\nu}}$ and $\nu > 0$ determines the asymmetry of the curve. Notice that, by choosing $Q=\nu=1$, $x_0$ corresponds to the point at which the curve is at its midpoint and has the maximum slope.

By using the generalized logistic function, with $Q=\nu=1$, to express the probability of success of a single attack $P_s(x)$, (sometimes $P_s$, for notation simplicity, in the following) we have
\begin{equation}
    P_s(x)=A+\frac{K-A}{1+e^{-B(x-x_0)}}. 
    \label{eq:genlog2}
\end{equation}
By increasing $x_0$, $P_s(x)$ shifts toward right. So, we associate the value of the complexity of the organization to $x_0$, to take into account that, for the same value of $x$, more complex organizations are expected to face attacks with a larger probability of success.

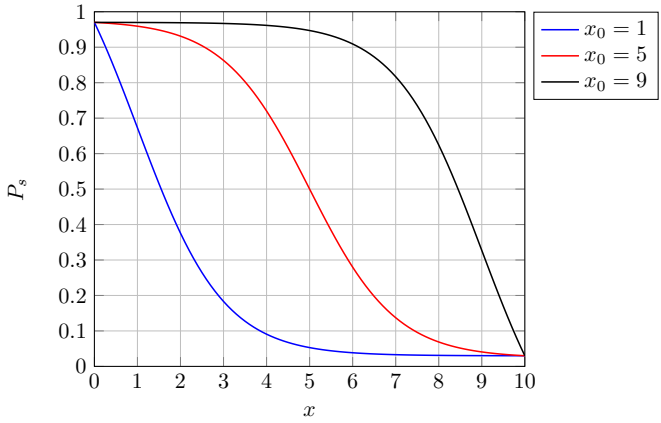
\begin{figure}[tb]
    \begin{centering}
    \input{Figures/varia_x0}
    \caption{Probability of success of a single attack ($P_s$) for different values of $x_0$. $B=-1$, $U=0.97$, $L=0.03$.}
    \label{fig:varia_x0}
    \end{centering}
\end{figure}

According to the discussion in Section \ref{subsec:maturity}, the maturity $x$ takes values in the range from $0$ to $10$. Moreover, we assume that the probability of success cannot reach the extreme and thus unrealistic values of $1$ and $0$. So, we set the maximum value $U$ at $x = 0$ and the minimum value $L$ at $x = 10$. Consequently, $K$ and $A$ depend on $x_0$ and can be easily obtained by solving the system with $f(0)= U$ and $f(10)= L$.
An example of curves for different values of $x_0$ is shown in Fig. \ref{fig:varia_x0}. 

Indeed, as mentioned above, a more significant value of the probability of success can be obtained by taking into account the maturity of the attackers, the latter playing the role of weight coefficient $w$.  Explicitly this means that the final value of the probability of success of a single attack results in
\begin{equation}
p^*=p_s(x)=wP_s(x).
\label{eq:pattack}
\end{equation}
A possible choice for the values of $w$ is given in Table \ref{table:w&nattra}, showing how more attractive organizations suffer attacks which are more likely to be successful because of the higher maturity of the attackers. In this case, we assume that a linear law models well the expected behaviour. We observe that, in case of non-malicious threats, $w$ is always equal to $1$, since attractiveness does not play any role.

\begin{table}[ht!]
\centering
   \caption{Possible values for $w$ as a function of the attractiveness of the organization.}
\begin{tabular}{cccccc}
        \hline
         Attractiveness & Very Low & Low & Medium & High & Very High \\
         \hline
          $w$ & 0.6 & 0.7 & 0.8 & 0.9 & 1 \\
         \hline
    \end{tabular}
    \label{table:w&nattra}
\end{table}

\subsection{Number of attacks}\label{subsec:noth}

Let us consider a time period containing $t$ time slots, each of duration $\Delta t$.
An attacker (malicious or non-malicious, according to the discussion in Section \ref{subsec:prothr})  may perform an attack attempt or not in any time slot, with a certain probability. Therefore, at most $t$ attack attempts will be suffered by the organization in the considered time period. Let us define two discrete random variables:
\begin{itemize}
    \item $\mathcal A$: represents the outcome of the attacker choice in a time slot. Either he performs an attack attempt ($\mathcal A=1$), or he does not ($\mathcal A=0$). Therefore, $\mathcal A$ is a binary random variable and follows a Bernoulli distribution. 
    \item $\mathcal N$: represents the number of attack attempts suffered by the organization in the considered time period. Its realization $n$ respects the condition $0\leq n\leq t$. 
\end{itemize}

We assume that $\mathcal N$ follows a  binomial distribution\footnote{An alternative, which might be considered when $n_{\mathrm{avg}}$ is significantly smaller than $t$, is the Poisson distribution, whose probability mass function is $
 \mathrm{Pr}(\mathcal N=n)=\frac{\lambda^n}{n!} e^{-\lambda}$, where $\lambda=n_{\mathrm{avg}}$.}. The PMF  (probability mass function) associated to the binomial distribution is therefore
\begin{align}
\mathrm{Pr}(\mathcal N=n)&\nonumber=\binom{t}{n}\mathrm{Pr}(\mathcal A=1)^n\mathrm{Pr}(\mathcal A=0)^{(t-n)}\nonumber \\ &=\binom{t}{n}\Bigg(\frac{n_{\mathrm{avg}}}{t}\Bigg)^n\Bigg(1-\frac{n_{\mathrm{avg}}}{t}\Bigg)^{(t-n)}. \label{eq:binomTN}
\end{align}

In \eqref{eq:binomTN} we have denoted by $n_{\mathrm{avg}}$ the average number of attack attempts suffered by the organization in the considered time period, so that, reminding the properties of the binomial distribution,  $\mathrm{Pr}(\mathcal A=1)=\frac{n_{\mathrm{avg}}}{t}=1-\mathrm{Pr}(\mathcal A=0)$. A possible choice is to pick $n_{\mathrm{avg}}$ as the number of attack attempts suffered by the organization in a previous time period having the same length of the considered one. 

Then, wishing to calculate the probability that the organization suffers at most (or at least) $n$ attack attempts, we have
\[
 \mathrm{Pr}(\mathcal N\leq n)=\sum_{k=0}^{n}  \mathrm{Pr}(\mathcal N=n)
 \]
 \[\mathrm{Pr}(\mathcal N\geq n)=\sum_{k=n}^{t}  \mathrm{Pr}(\mathcal N=n),
\]
respectively.

\subsection{Likelihood of occurrence of cyber incidents}

In the previous section, we have analyzed the probability that the organization suffers a certain number of attack attempts. Even though this represents an interesting information for the stakeholders, it is more relevant to estimate the likelihood of these attacks being successful. In fact, in the definition of risk \cite{sp800-30} the likelihood of a successful attack, i.e., a cyber incident, is multiplied by its impact, and the impact of non-successful attacks is obviously $0$. We therefore introduce a further random variable $\mathcal{S}$, which models the number of cyber incidents faced by the organization in the considered period of time. We consider two scenarios:
\begin{itemize}
    \item the likelihood of facing $s$ cyber incidents, after at most $t$ attempts by the attacker. In this simplistic approach, it is assumed that the organization does not promptly realize when it is successfully attacked, and, therefore, does not take the appropriate countermeasure to change posture; 
		\item the likelihood of facing exactly one cyber incident, after at most $t$ attack attempts by the attacker. In this more realistic approach, it is assumed that the organization immediately notices the breach, and tries to improve its posture.
\end{itemize}

\subsubsection{Organizations which do not change posture}\label{subsubsec:ncp}

In this scenario, having assumed that the organization does not react immediately after experiencing one or more cyber incidents, we can model the attacks as Bernoulli experiments, with success probability given by \eqref{eq:pattack}. We can actually assume that also the outcome of the single attack attempt is a continuous random variable $\mathcal{P}$ with realization $p$. In particular, taking into account the uncertainty inherent in the problem (where the maturity index, for example, results from the opinion of some experts or from the assessment of questionnaires) we assume that $\mathcal{P}$ follows a PERT (Project Evaluation and Review Techniques) distribution, with  $p_m=p_s(\min\{x+q,10\})$,  $p_M=p_s(\max\{x-q,0\})$, for some arbitrary value of $q$, and $p^* = p_s(x)$ respectively the minimum, maximum, and most likely values. In the following, we have considered $q=1$. Therefore, the  probability that $s$ out of $n\leq t$ trials are successful is given by
\[
\mathrm{Pr}(\mathcal{S}= s|\mathcal{N}=n)=\int_{p_m}^{p_M}\binom{n}{s}p^s(1-p)^{(n-s)}f_{\mathcal{P}}(p)\mathrm{d}p,
\]
where
\[
f_{\mathcal{P}}(p)=\frac{(p-p_m)^{\alpha-1}(p_M-p)^{\beta-1}}{B(\alpha,\beta)(p_M-p_m)^{\alpha+\beta-1}},
\]
being $\alpha=1+4\frac{p^*-p_m}{p_M-p_m}$ and $\beta=1+4\frac{p_M-p^*}{p_M-p_m}$, while $B(\alpha,\beta)=\int_{0}^{1} x^{\alpha-1}(1-x)^{\beta-1} dx$ is the Beta function. For the likelihood, we finally obtain
\begin{align}
L^{(\mathrm{NC})}(s)&=\mathrm{Pr}(\mathcal{S}=s)\nonumber\\&=\sum_{n=1}^{t} \mathrm{Pr}(\mathcal{S}=s|\mathcal{N}=n) \mathrm{Pr}(\mathcal{N}=n). 
\label{eq:teval}
\end{align}

\subsubsection{Organizations which change posture}\label{subsubsec:cp}

In this scenario, we assume that the organization runs into a single cyber incident, detects it immediately and decides to promptly take countermeasures, changing its posture. Therefore, a new cyber risk assessment should be performed after one successful attack, since the initial conditions, and the value of the maturity index above all, should have been changed. In this case, we consider a cumulative geometric distribution, \textit{i.e.},
\[
\mathrm{Pr}(\mathcal{S}= 1|\mathcal{N}=n)=\int_{p_m}^{p_M}\sum_{k=1}^{n}p(1-p)^{(k-1)}f_{\mathcal{P}}(p)\mathrm{d}p.
\]
Then, the likelihood of occurrence of a single cyber incident can be obtained by considering all possible values of $n$, \textit{i.e.},
\begin{equation}
L^{(\mathrm{C})}=\mathrm{Pr}(\mathcal{S}=1)=\sum_{n=1}^{t} \mathrm{Pr}(\mathcal{S}=1|\mathcal{N}=n) \mathrm{Pr}(\mathcal{N}=n).
\label{eq:likeligeom}
\end{equation}


\section{Cyber incident model in HTMA and FAIR}\label{sec:merging}
In this section we show how the values of the likelihood (or frequency) computed in Sections \ref{subsubsec:ncp} and \ref{subsubsec:cp}  can be plugged into HTMA and FAIR, minimizing the subjectivity of the assessor and, rather, allowing the computation of risk according to the posture of the considered organization.

\subsection{Use of MAGIC with HTMA}

According to the description in Section \ref{subsec:accatma}, HTMA does not take into account the possibility that a threat happens more than once in a year  and, therefore, we must consider the likelihood that the organization faces the threat just once. This means that the scenario described in Section \ref{subsubsec:cp} must be applied. As for the monetary impact of these threats, they are not object of investigation in this paper and, therefore, we rely on available results.

In particular, MAGIC can be used to estimate the likelihood of occurrence of the  threats considered in the Monte Carlo simulation (point 1. in Section \ref{subsec:accatma}). In fact, instead of relying on the expertise of the assessor, it is possible to employ the likelihood obtained through our cyber incident model  \eqref{eq:likeligeom} as input to the HTMA method. However, we need to associate a likelihood to every single cyber threat in the considered list (and not a single, general likelihood value) and, therefore, the model must be adapted. Explicitly the procedure works as follows. We start by defining a list of threats. This list is then combined with the list of considered controls through a table containing weight coefficients $\omega_{i,j}$ as shown in Table \ref{table:wecon}, where $\tau_j$ and $\gamma_i$ denote the $j$-th threat and the $i$-th control, respectively.

\begin{table}[ht!]
\centering
   \caption{Weight coefficients for likelihood assessment in HTMA.}
\begin{tabular}{c|ccccc}
         & $\tau_1$ & $\tau_2$ & $\cdots$ & $\tau_n$ \\
         \hline
          $\gamma_1$ & $\omega_{1,1}$ & $\omega_{1,2}$ &$\cdots$ & $\omega_{1,n}$\\
          $\gamma_2$ & $\omega_{2,1}$& $\omega_{2,2}$ & $\cdots$& $\omega_{2,n}$ \\
          $\vdots$ &  $\vdots$&  $\vdots$& $\ddots$ & $\vdots$  \\
          $\gamma_m$ & $\omega_{m,1}$ & $\omega_{m,1}$ & $\cdots$&  $\omega_{m,n}$ \\
    \end{tabular}
    \label{table:wecon}
\end{table}

First of all, one should answer the question ``if the control $\gamma_j$ is not implemented, does the risk relative to the threat $\tau_i$ increase?''. If not, then $\omega_{i,j}$ is $0$, otherwise a non-zero value can be assigned to $\omega_{i,j}$, according to some predetermined rule. This way, for each threat we can select a subset of controls that includes only those controls that have a non-zero weight $\omega$ for that specific threat. Notice that the subsets are not necessarily disjoint. We can compute a maturity index for each of these subsets and obtain the likelihood of occurrence of a cyber incident caused by the threat $\tau_j$, noted by $L_j$, following the procedure presented in the previous sections.

\subsection{Use of MAGIC with FAIR}

Referring to the FAIR ontology in Fig. \ref{fig:schemeFAIR}, the blocks that are directly influenced by our model are the TEF, the Vulnerability and the LEF (both for the primary and the secondary threat event). Clearly, these blocks will also indirectly influence the upper layers of the ontology  which, however, also depend on blocks which are not object of our analysis. We remind that the TEF is defined as the frequency with which, in a given time period, the attacker tries to breach the organization. In Section \ref{subsec:noth}, we have devised a model that returns the probability that the organization faces a given number of attacks. As discussed below, the link between number of attacks and frequency of the attacks is immediate, once the reference time period has been defined. The Vulnerability, defined as the probability of success of any of these breaches, assumes the same meaning of the probability of success of the attacks, defined in Section \ref{subsec:prothr} and denoted as $p^*$.  Finally, the LEF is obtained by combining the other two parameters; it is defined as the frequency with which the attacker succeeds, causing a monetary loss to the organization. Also this parameter has been covered by our analysis, especially that in Section \ref{subsubsec:ncp}, where the likelihood that the organization is breached a given number of times is computed through \eqref{eq:teval}. Also in this case, the connection between likelihood and frequency is straightforward. 

In the original FAIR approach, it is assumed that the LEF follows a PERT distribution, whose parameters must be computed based on experts' estimates. Using MAGIC, we maintain the Monte Carlo approach, as already done for the HTMA approach, but we sample the LEF from the distribution of $\mathcal{S}$ (\textit{i.e.}, $\mathrm{Pr}(\mathcal{S}=s)$) reported in \eqref{eq:teval}, which is computed taking into account both the TEF and the Vulnerability. The random sampling can be performed very easily, since it is referred to a discrete distribution: we can associate a probability to each allowed value of $s$ and sample according to these probabilities. Notice that, once $s$ is obtained by random sampling, the LEF in the considered framework can be simply computed as $\mathrm{LEF}=\frac{s}{t}$.

\section{Numerical results}\label{sec:Results}
In this section we provide some numerical results, which aim to validate the effectiveness of MAGIC.

\subsection{Case study with HTMA}
\label{subsec:HTMA}

In order to take into account a realistic scenario, we have considered the list of threats given in \cite[Table 1]{santini2019} (originally taken from \cite{ponemon}), and here reported, for the sake of clarity, in Table \ref{table:threats}. Furthermore, as a case study, we have analyzed a dummy organization in the ``Healthcare'' sector. We have simulated a posture assessment, finding, for each threat, a maturity index as in Table \ref{table:threats}. As for the attractiveness, according to the discussion in Section \ref{subsec:attractiveness}, we have found that organizations in the healthcare sector are very highly attractive. Indeed, this is not surprising, and well-known in the literature. Finally, we have considered a range of values for the complexity index, from $4.5$ to $7.5$ with step $0.5$. We have also considered different values of $n_{\mathrm{avg}}$, ranging from $2$ to $5$; we have assumed that, in each simulation, $n_{\mathrm{avg}}$ is the same for all threats (for the sake of comparison between different threats). As in \cite{RafaianiAEIT}, we choose $B=-2$, $U=0.97$ and $L=0.03$ as parameters of the generalized logistic function. Looking at Table \ref{table:threats} we notice that we are talking about an attractive organization, which however has some serious maturity shortcomings, which may lead to weaknesses against cyber threats.

\begin{table*}[ht!]
\centering
   \caption{Parameters of the proposed model, as inputs of HTMA for the considered scenario, when the complexity index is $5$ and $n_{\mathrm{avg}}=4$.}
\begin{tabular}{ccccccc}
        \hline
         ID & Threat & Maturity index&$p^*$&$p_m$&$p_M$&$L^{\mathrm{C}}$\\
         \hline
          1& Malware & $4.3$& $0.50$ & $0.28$ & $0.72$ &$0.86$\\
2 &Web-based attacks & $5.6$ &$0.23$ & $0.11$ & $0.43$&$0.61$\\
3 &Denial of services & $3.6$ &$0.66$ & $0.43$& $0.83$&$0.92$\\
4 &Malicious insiders & $1.9$ &$0.90$& $0.79$& $0.95$&$0.97$\\
5 &Phishing and social engineering& $3.6$&$0.66$& $0.43$& $0.83$&$0.92$\\
6 &Malicious code & $6.0$& $0.17$ & $0.08$ & $0.34$&$0.52$\\
7 &Stolen devices & $4.8$&$0.38$& $0.19$ & $0.62$&$0.78$\\
8 &Ransomware & $5.1$&$0.32$& $0.16$ & $0.55$&$0.72$ \\
9 &Botnets & $4.3$ &$0.50$& $0.28$ & $0.72$& $0.86$ \\
         \hline
    \end{tabular}
    \label{table:threats}
\end{table*}

The parameters of the PERT distribution, along with the likelihood of occurrence of each threat in a year ($t=365$ and $\Delta t = 1$, assuming that the attacker performs at most one attack a day), computed by \eqref{eq:likeligeom}, are also shown in Table \ref{table:threats} for $n_{avg}=4$ and complexity index $5$. As expected, due to the relatively low values of the maturity indexes associated to each threat, the organization has a relatively high probability of suffering a cyber incident through them. Still, threats corresponding to lower maturity indexes are less likely to be suffered than threats corresponding to higher maturity indexes.

Regarding the monetary impacts, we have considered those in \cite[Table 2]{santini2019} and also reported in Table \ref{tab:impatti_r}, for the sake of completeness. In order to keep a uniform notation with the rest of the paper, we have applied a currency exchange from dollars (used in \cite{santini2019}) to Euro. Notice that, assuming that all threats happen and the associated cyber incidents cause the largest possible loss (given by the upper value of the corresponding interval), we obtain an upper bound for the total loss which, in the proposed example, is a value between $11.5$ and $12$ million Euro (precisely, $11.8361$ million Euro). 

\begin{table}[ht!]
\centering
   \caption{Impact range of the considered threats, in million Euro.}
\begin{tabular}{ccccccc}
        \hline
         ID & Threat & Impact range\\
         \hline
          1& Malware & $[2.1360,2.3941]$ \\
2 &Web-based attacks & $[1.8156,2.0381]$\\
3 &Denial of services &  $[1.4151,1.5842]$\\
4 &Malicious insiders & $[1.2816,1.4329]$\\
5 &Phishing and social engineering&  $[1.1748,1.3172]$ \\
6 &Malicious code & $[1.1659,1.2994]$\\
7 &Stolen devices &  $[0.77875,0.87576]$\\
8 &Ransomware &   $[0.48060,0.53845]$\\
9 &Botnets &  $[0.31684,0.35600]$\\
         \hline
    \end{tabular}
    \label{tab:impatti_r}
\end{table}

As usual, we express the final output of the model as a LEC, summarizing the probability that the loss faced by the organization will be greater than or equal to a certain value, obtained with a Monte Carlo simulation with $10,000$ trials. 

\begin{figure}[tb]
    \begin{centering}
    \includegraphics[width=0.48\textwidth]{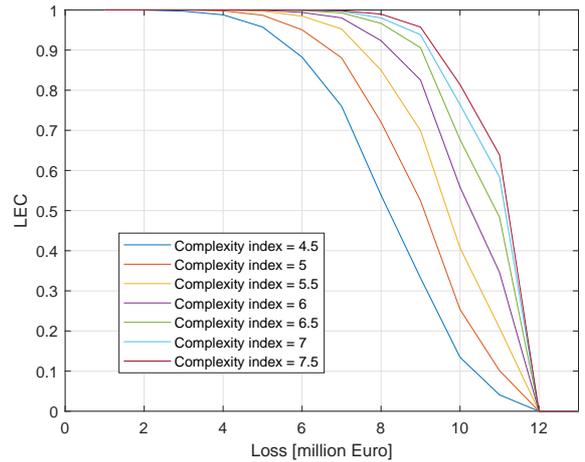}
    \caption{LEC for the considered (very highly attractive) organization, for different values of the complexity index and $n_\mathrm{avg}=4$, using the HTMA method.}
    \label{fig:HTMA_MC}
    \end{centering}
\end{figure}

We show the LECs, obtained using the aforementioned values of the complexity and considering $n_\mathrm{avg}=4$, in Fig. \ref{fig:HTMA_MC}. Notice that, coherent with our analysis, more complex organizations are more likely to suffer cyber attacks and, therefore, they will be subject to larger monetary losses. 

\begin{figure}[tb]
    \begin{centering}
    \includegraphics[width=0.48\textwidth]{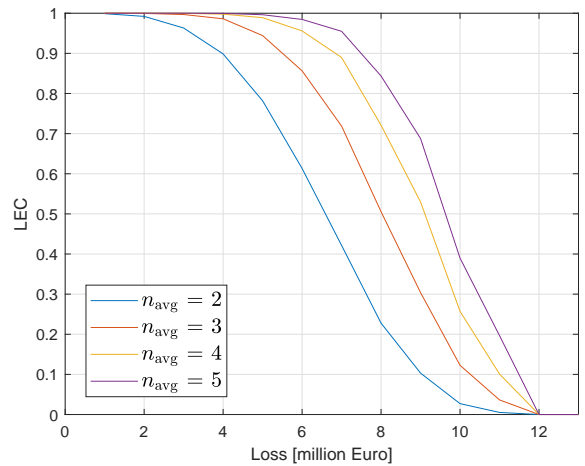}
    \caption{LEC for the considered (very highly attractive) organization, for different values of $n_\mathrm{avg}$ and complexity index equal to $5$, using the HTMA method.}
    \label{fig:HTMA_MC_n}
    \end{centering}
\end{figure}

Finally, we have shown in Fig. \ref{fig:HTMA_MC_n} the LECs for complexity index equal to $5$ and different values of $n_{\mathrm{avg}}$. Also in this case, as expected, the losses increase for increasing values of $n_{avg}$, which we have chosen as the number of attack attempts suffered in the previous year, relatively to each threat.

\subsection{Case study with FAIR}\label{subsec:fairres}

We have considered the same scenario described in \cite[Chapter 8]{fair}. For the sake of simplicity, we have assumed that the Loss Magnitude is only caused by primary losses and then ignored secondary risk. Clearly, this assumption can be removed without changing the rationale of the model. Concerning the Loss Magnitude, we have not changed the minimum, maximum, and most likely values (and neither we have changed the confidence level) for the considered categories  in \cite[Chapter 8]{fair}: productivity, response, replacement, fines \& judgements, reputation and competitive advantage. The only categories  having non-zero impact values are response and replacement and, for the sake of completeness, we have reported them in Table \ref{tab:impatti}.  Also in this case, we have considered a dummy organization in the ``Healthcare'' sector. Simulating a different posture assessment from the previous section, we have found a complexity index equal to  $5.2$ and a maturity index equal to $6.9$. Being in the healthcare sector, the organization is considered again very highly attractive.  The values of $B$, $U$ and $L$ are once again set as $-2$, $0.97$ and $0.03$, respectively. The analysis is carried out on daily intervals within a year, \textit{i.e.}, $t=365$ and $\Delta t= 1$, and we have considered  $n_\mathrm{avg}=84$. Notice that this value is significantly larger than those adopted for the examples in Section \ref{subsec:HTMA}, just in view of enlarging the ensemble of considered scenarios.

\begin{table*}[ht!]
\centering
   \caption{Impact range of the considered event, in €.}
\begin{tabular}{ccccccc}
        \hline
         Category & Minimum & Maximum & Most Likely & Confidence \\
         \hline
          Response & $2,750$& $8,250$  &$22,000$ & $20$ \\
          Replacement & $20,000$& $30,000$  &$50,000$ & $20$ \\
         \hline
    \end{tabular}
    \label{tab:impatti}
\end{table*}

\begin{figure}[tb]
    \begin{centering}
    \includegraphics[width=0.48\textwidth]{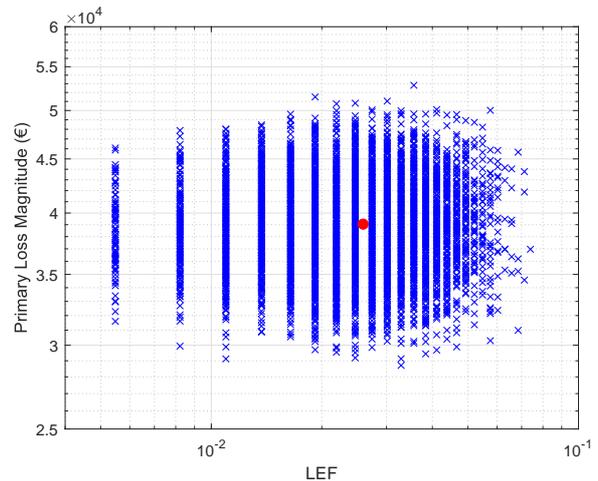}
    \caption{Loss magnitude for the considered organization, which is very highly attractive, when the complexity index is $5.2$ and the maturity index is $6.9$, obtained through the FAIR method.}
    \label{fig:FAIR_MC}
    \end{centering}
\end{figure}

Running the Monte Carlo simulation, we obtain the Loss Magnitudes in Fig. \ref{fig:FAIR_MC}. Each point corresponds to a different simulation.  The red circle in the figure represents the mean value (for both axes). Defining the $k$-th  percentile as the score below which the $k\%$ of the scores fall in the given distribution, we have found that the $10$-th percentile and the $90$-th percentile, for this example, are approximately $18,000$ € and $58,000$ €, respectively. The results of the Monte Carlo simulation are also summarized in Table \ref{tab:sumMonte} where, besides the minimum, the maximum and the mean, also the mode  (that is the most frequent value occurred in the Monte Carlo simulation, differently from the most likely value, that is the result of the statistical inference for the same quantity) is reported.

\begin{table*}[ht!]
\centering
   \caption{Summary of the results of the Monte Carlo simulation, when the complexity index is $5.2$ and the maturity index is $6.9$.}
\begin{tabular}{ccccccc}
        \hline
         & Minimum & Mean & Mode & Maximum   \\
         \hline
          Primary Loss Events per year  & 1 & 9.5 & 8 & 27\\
          Primary Loss Magnitude (€) & $28,709$ & $39,045$ & $32,948$ &$52,822$\\
     
         Total Loss Exposure (€) &$35,369$&$369,260$&$295,520$&$1,141,600$\\ \hline
    \end{tabular}
    \label{tab:sumMonte}
\end{table*}

Clearly, fixed $t$, a smaller value of $\Delta t$  increases the number of possible attacks per year (which is now $365$), thus increasing the number of possible values of the LEF, in its turn.

\begin{figure}[tb]
    \begin{centering}
    \includegraphics[width=0.48\textwidth]{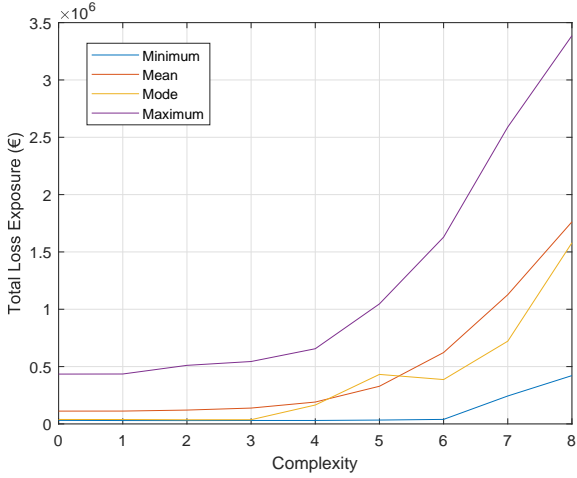}
    \caption{Minimum, mode, mean and maximum of the total loss exposure, for different values of the complexity index.}
    \label{fig:changecomple}
    \end{centering}
\end{figure}

\begin{figure}[tb]
    \begin{centering}
    \includegraphics[width=0.48\textwidth]{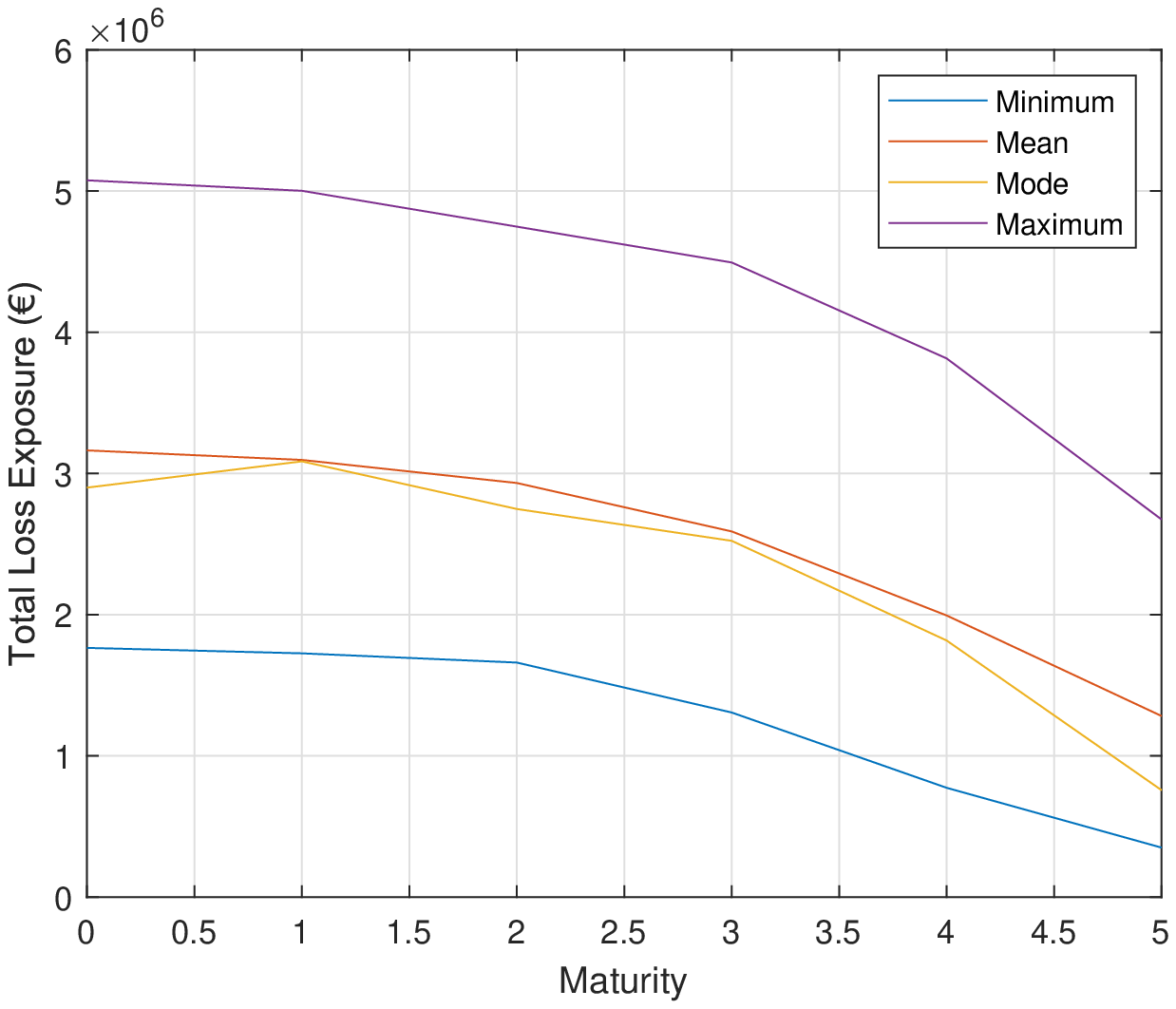}
    \caption{Minimum, mode, mean and maximum of the total loss exposure, for different values of the maturity index.}
    \label{fig:changematurity}
    \end{centering}
\end{figure}

\begin{figure}[tb]
    \begin{centering}
    \includegraphics[width=0.48\textwidth]{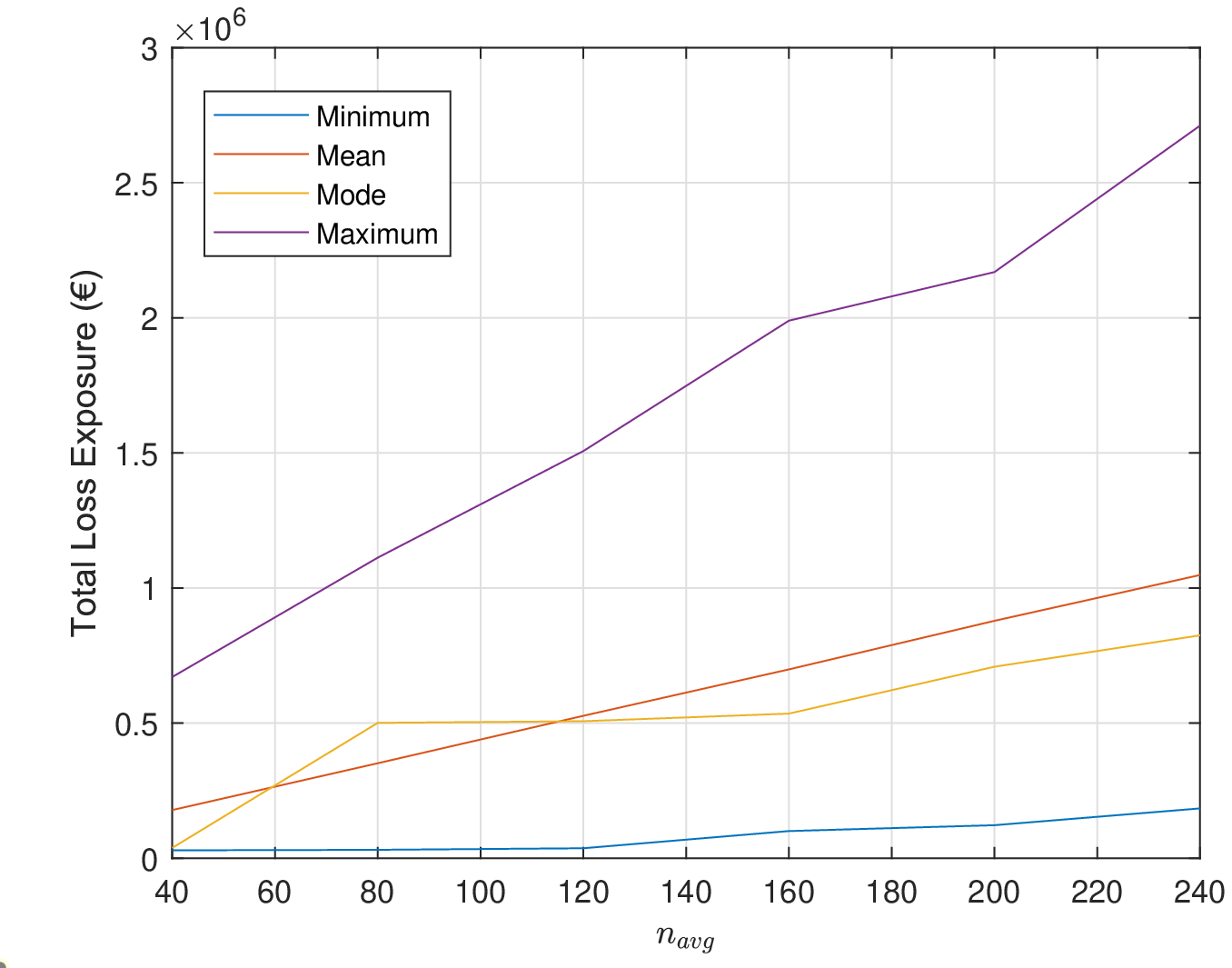}
    \caption{Minimum, mode, mean and maximum of the total loss exposure, for different values of $n_\mathrm{avg}$.}
    \label{fig:changenavg}
    \end{centering}
\end{figure}

In order to evaluate the impact of different values of the maturity index, complexity index and $n_{\mathrm{avg}}$, we have fixed two of these parameters and let the other change, obtaining Figs. \ref{fig:changecomple}, \ref{fig:changematurity} and \ref{fig:changenavg}, respectively. The results are coherent with our analysis: on the one hand, for increasing values of the complexity and of $n_\mathrm{avg}$, the total loss exposure increases; on the other hand, the total loss exposure decreases when the organization has increasing values of the maturity.

\section{Comparison and discussion}\label{sec:comdiscuss}

In this section we perform both a quantitative and a qualitative comparison of MAGIC with other methods.

\subsection{Quantitative comparison}\label{subsec:quanticomp}

Let us consider the same organization in the ``Healthcare" sector as in Section \ref{subsec:HTMA}, for which a posture assessment has returned a complexity index equal to $5$. The maturity index associated to each of the considered threats is given in Table \ref{table:threats} (third column).
Owing to its sector, the organization is considered very highly attractive.

We have computed the likelihood of occurrence of cyber incidents caused by the same threats using the basic version of CVSS-based risk assessment method, relying on the fundamentals described in \cite{cvss}. The low level metrics we use for the comparison are those given in \cite{cvss}, which we also report in Table \ref{tab:CVSS} for the sake of completeness. The assignment of the numerical values to these CVSS low level metrics, which is the possibly subjective phase of this approach, has been performed by means of a brainstorming session, based on the experience of the authors and on available evidences. It must be stressed that some assignments are not subjective: for example, the Access Vector of stolen devices or malicious insiders is objectively ``Local''. However, this is neither true for all metrics, nor for all threats.

\begin{table*}[ht!]
\centering
   \caption{Low level metrics for likelihood assessment using CVSS.}
\begin{tabular}{cccc}
        \hline
         Metric & CVSS & Metric evaluation & Numerical value   \\
         \hline
          \multirow{3}{*}{Access Vector (AV)}  & \multirow{3}{*}{Base} & Local & $0.4$\\
          &&Adjacent&$0.6$\\
          &&Network&$1$\\
          \hline
                    \multirow{3}{*}{Authentication (Au)}  & \multirow{3}{*}{Base} & Multiple & $0.5$\\
          &&Single&$0.55$\\
          &&None&$1$\\
          \hline
          
          \multirow{3}{*}{Access Complexity (AC)}  & \multirow{3}{*}{Base} & High & $0.5$\\
          &&Medium&$0.75$\\
          &&Low&$1$\\
          \hline
          
           \multirow{5}{*}{Exploitability (E) }  & \multirow{5}{*}{Temporal} & Unproven & $0.85$\\
           &&Proof of concept&$0.9$\\
          &&Functional&$0.95$\\
           &&High&$1$\\
           &&Not defined&$1$\\
          \hline

\multirow{4}{*}{Report Confidence (RC)}  & \multirow{4}{*}{Temporal} & Unconfirmed & $0.9$\\
          &&Uncorroborated&$0.9$\\
          &&Confirmed&$1$\\
           &&Not defined&$1$\\
          \hline    \end{tabular}
    \label{tab:CVSS}
\end{table*}

Lastly, a team of certified external experts, taking as inputs the status of the organization used for the posture assessment, in order to find the maturity, complexity and attractiveness indexes, has provided estimates for the likelihood of occurrence of cyber incidents deriving from the nine considered threats. This expert-based approach is used in many modern cyber-risk assessment methods, such as HTMA, FAIR, MAGERIT, CRAMM, and many others.

The results obtained through MAGIC and the aforementioned approaches are compared in Table \ref{table:threats_expert}, where we denote the likelihoods resulting from the newly proposed method, the CVSS-based method and the expert estimates as $L^{\mathrm{C}}$,  $L^{\mathrm{CVSS}}$ and  $L^{\mathrm{EXPERT}}$, respectively.

\begin{table*}[ht!]
\centering
   \caption{Likelihood of occurrence of the considered threats, obtained using different methods.}
\begin{tabular}{ccccc}
        \hline
         ID & Threat & $L^{\mathrm{C}}$ &  $L^{\mathrm{CVSS}}$&  $L^{\mathrm{EXPERT}}$\\
         \hline
          1& Malware & $0.86$&$0.90$&$0.80$\\
2 &Web-based attacks &$0.61$&$0.64$& $0.75$\\
3 &Denial of services & $0.92$&$0.95$&$0.90$\\
4 &Malicious insiders & $0.97$&$0.15$&$0.55$\\
5 &Phishing and social engineering&$0.92$&$0.90$&$0.95$\\
6 &Malicious code & $0.52$&$0.81$&$0.70$\\
7 &Stolen devices &$0.78$&$0.09$&$0.45$\\
8 &Ransomware &$0.72$ &$0.99$&$0.85$\\
9 &Botnets & $0.86$&$0.86$& $0.80$\\
         \hline
    \end{tabular}
    \label{table:threats_expert}
\end{table*}

We notice that, in several cases,  the three considered methods provide comparable results. However, in some cases, the numerical methods (MAGIC and the CVSS-based) give contrasting results. We argue this is due to the incapability of the CVSS-based method to catch that different organizations may react differently to the same threats, if the state of their organizational infrastructure is dissimilar. In fact, the CVSS-based method considers only marginally and indirectly the degree of compliance of the organization to the best practices for threat prevention, whereas MAGIC takes it into account in the preliminary posture assessment. In contrast, on the one hand, the external experts evaluation seems very balanced in most situations but, on the other hand, as already asserted, it is subjective, in that a different team of experts might return significantly different results. Moreover, the involvement of external experts is costly, it is time consuming, and requires a significant amount of data to be processed, which may not be available.

\subsection{Qualitative comparison}

In this section we provide a qualitative comparison of several cyber risk assessment and/or management methods. The considered approaches, the evaluated metrics and the results of the assessment are shown in Table \ref{tab:my_qualcomp}.

\begin{table*}[ht]
    \centering
    \caption{Qualitative comparison between the  considered methods. Legend: RM=Risk Management; RA=Risk Assessment; LA=Likelihood Assessment.}
    \begin{threeparttable}
    \begin{tabular}{l|>{\centering\arraybackslash}m{1.5cm}p{3.5cm}m{4cm}c}
    \hline
    Method & Scope & Likelihood Assessment & Assessors & Target Organizations \\ \hline
    ISO/IEC 27005:2018 & RM & Guidelines only & External experts  & All  \\
    NIST SP 800-30 & RA & Guidelines only & External experts & Governmental\tnote{*}  \\
    EBIOS & RM & Qualitative & Team of different stakeholders & All \\
    CRAMM & RA & Qualitative  & External experts & Large scale\tnote{*}  \\
    OCTAVE & RM & Qualitative & Internal experts & All \\
    MEHARI & RM & Qualitative & Internal experts & All \\
    MAGERIT & RM & Numerical & Calibrated experts & Public Administration\tnote{*} \\
    IRAM2 & RM & Qualitative & Internal experts & All \\
    IT-Grundschutz & RM &  Qualitative & Internal and external experts & All \\
    CORAS & RA & Qualitative or numerical & Internal and external experts & Medium and large scale  \\
    HTMA & RA & Numerical  & Calibrated experts & Medium and large scale \\
    FAIR & RA & Numerical &Calibrated  experts & Medium and large scale  \\
    LiSRA & RA & Numerical & External experts & All \\
    Fuzzy-logic based & LA &Numerical & External experts & Medium and large scale\\
    CVSS-based & LA &Numerical & Internal experts & All\\
    \textbf{MAGIC} & \textbf{LA} & \textbf{Numerical} & \textbf{Internal experts}& \textbf{All} \\ \hline
    \end{tabular}
    \begin{tablenotes}\footnotesize
\item[*]Can be extended to other types of organizations.
\end{tablenotes}
  \end{threeparttable}

    \label{tab:my_qualcomp}
\end{table*}

In short, we notice that among the most relevant existing risk assessment/management methods, most of them require the subjective estimates of calibrated or external experts. Clearly, the extensive use of external experts (described in Section \ref{subsec:quanticomp}) restricts the range of the organizations that can afford it to those in the medium and large scale, making the associated methods barely applicable to small organizations. 
Some methods, such as MEHARI or MAGERIT, rely on internal experts, like our method, but they implement only qualitative approaches. Therefore, they do not provide a numerical estimate of the likelihood of occurrence of cyber incidents, leading to results which might need an involved interpretation and may not be easily reproducible. Even in the CVSS-based methods, where the likelihood is numerically estimated based on internal experts analyses, the process is still subjective, since there is no automatic and universal way to assign scores to the threats exploiting the considered vulnerabilities. Moreover, none of the considered methods directly relate the likelihood of occurrence of cyber incidents to the posture of the organization. MAGIC tries to overcome all these shortcomings, by providing a numerical approach for the likelihood assessment, relying on the simple initial task of some internal experts compiling questionnaires about the status of the target organization.

\section{Conclusion}\label{sec:concl}

Most existing cyber risk assessment methods are affected by subjectivity or require a high level of expertise from the organizations, which makes them barely practical in real-case scenarios. We have tackled this issue by proposing a novel probabilistic method, called MAGIC that, based on the organization posture, estimates the likelihood (or the frequency) of occurrence of a cyber incident or, more generally, of a list of cyber incidents. Through numerical simulations we have shown how MAGIC can be plugged into statistical cyber risk assessment methods, like HTMA and FAIR, to eventually assess the risk as the likelihood of occurrence of cyber incidents combined with their impact.  We have performed both qualitative and quantitative comparisons with alternative approaches, showing that our method is reliable and general, in the sense that it can provide inputs to other risk assessment and/or risk management methods. Moreover, MAGIC tries to catch all the advantages of existing likelihood assessment methods. In conclusion, due to its simplicity and rather low computational demand, we argue that MAGIC can be applied to any type of organization (small, medium, or large), without any loss of generality.

\section*{Acknowledgment}

The authors are very grateful to Giovanni Libertini and to the team of cybersecurity analysts of Ancharia S.r.l. for their precious insights.  

\bibliographystyle{IEEEtran}
\bibliography{ref.bib}
\end{document}

%% file: Figures/fair.tex
\resizebox{1.8\columnwidth}{!}{
\tikzstyle{block} = [rectangle, draw, fill=gray!20, text width=3.8cm, text centered, rounded corners, minimum height=1.6cm, minimum width=4cm]
\tikzstyle{line} = [draw, -latex']
    
\begin{tikzpicture}
\newcommand{\blockdist}{4.4}
\tikzset{
myarrow/.style={-, >=latex', shorten >=1pt, thick}
}
    \node [block] (A1) at (0,0) {\Large{Contact Frequency}};
    \node [block] (A2) at (\blockdist,0) {\Large{Probability of Action}};
    \node [block] (A3) at (2*\blockdist,0) {\Large{Threat Capability}};
    \node [block] (A4) at (3*\blockdist,0) {\Large{Difficulty}};
    \node [block] (A5) at (6*\blockdist,0) {\Large{Secondary Loss Event Frequency}};
    \node [block] (A6) at (7*\blockdist,0) {\Large{Secondary Loss Magnitude}};
    
    \node [block] (B1) at (\blockdist/2,3) {\Large{Threat Event Frequency}};
    \node [block] (B2) at (2*\blockdist+\blockdist/2,3) {\Large{Vulnerability}};
    \node [block] (B3) at (4*\blockdist+\blockdist/2,3) {\Large{Primary Loss}};
    \node [block] (B4) at (6*\blockdist+\blockdist/2,3) {\Large{Secondary Risk}};
    
    \node [block] (C1) at (\blockdist+\blockdist/2,6) {\Large{Loss Event Frequency}};
    \node [block] (C2) at (5*\blockdist+\blockdist/2,6) {\Large{Loss Magnitude}};

    \node [block] (D1) at (3*\blockdist+\blockdist/2,9) {\Large{Risk}};

    \draw [myarrow] (A1.north) -- (0,1.5) -- (\blockdist/2,1.5) -- (B1.south);
    \draw [myarrow] (A2.north) -- (\blockdist,1.5) -- (\blockdist/2,1.5) -- (B1.south);

    \draw [myarrow] (A3.north) -- (2*\blockdist,1.5) -- (2*\blockdist+\blockdist/2,1.5) -- (B2.south);
    \draw [myarrow] (A4.north) -- (3*\blockdist,1.5) -- (2*\blockdist+\blockdist/2,1.5) -- (B2.south);
    
    \draw [myarrow] (A5.north) -- (6*\blockdist,1.5) -- (6*\blockdist+\blockdist/2,1.5) -- (B4.south);
    \draw [myarrow] (A6.north) -- (7*\blockdist,1.5) -- (6*\blockdist+\blockdist/2,1.5) -- (B4.south);
    
    \draw [myarrow] (B1.north) -- (\blockdist/2,4.5) -- (\blockdist+\blockdist/2,4.5) -- (C1.south);
    \draw [myarrow] (B2.north) -- (2*\blockdist+\blockdist/2,4.5) -- (\blockdist+\blockdist/2,4.5) -- (C1.south);
    
    \draw [myarrow] (B3.north) -- (4*\blockdist+\blockdist/2,4.5) -- (5*\blockdist+\blockdist/2,4.5) -- (C2.south);
    \draw [myarrow] (B4.north) -- (6*\blockdist+\blockdist/2,4.5) -- (5*\blockdist+\blockdist/2,4.5) -- (C2.south);
    
    \draw [myarrow] (C1.north) -- (\blockdist+\blockdist/2,7.5) -- (3*\blockdist+\blockdist/2,7.5) -- (D1.south);
    \draw [myarrow] (C2.north) -- (5*\blockdist+\blockdist/2,7.5) -- (3*\blockdist+\blockdist/2,7.5) -- (D1.south);

\end{tikzpicture}
}

%% file: Figures/model.tex
\resizebox{2\columnwidth}{!}{

\tikzstyle{line} = [draw, -latex']
    
\begin{tikzpicture}
\tikzset{
myarrow/.style={->, >=latex', shorten >=1pt, very thick}
}
\tikzset{
myarrow2/.style={->, >=latex', shorten >=1pt, very thick, dashed}
}
\tikzset{
myarrow3/.style={->, >=latex', shorten >=1pt, very thick, dotted}
}
    \node (mat_org) at (2.5,0) {\begin{tabular}{c} \textbf{\Large{Maturity}} \\ of the organization \end{tabular}};
    \node (compl_org) at (7.5,0) {\begin{tabular}{c} \textbf{\Large{Complexity}} \\ of the organization \end{tabular}};
    \node (mat_agents) at (0,3) {\begin{tabular}{c} \textbf{\Large{Maturity}} \\ of attackers \end{tabular}};
    \node (prob_success) at (5,3) {\begin{tabular}{c} \textbf{\Large{Probability}} \\ \textbf{\Large{of success}} \\ of an attack \end{tabular}};
    \node (awareness) at (-10,4.5) {\begin{tabular}{c} \textbf{\Large{Awareness}} \\ of the employees \end{tabular}};
    \node (attr_org) at (-5,4.5) {\begin{tabular}{c} \textbf{\Large{Attractiveness}} \\ of the organization \end{tabular}};
    \node (num_threats) at (0,6) {\begin{tabular}{c} \textbf{\Large{Number}} \\ of attacks \end{tabular}};
    \node (likelihood) at (5,6) {\begin{tabular}{c} \textbf{\Large{Likelihood}} \\ of occurrence of a \\ cyber incident \end{tabular}};
    \node (impact) at (10,6) {\begin{tabular}{c} \textbf{\Large{\textcolor{mybluegray}{Impact}}} \\ \textcolor{mybluegray}{derived from the occurrence} \\ \textcolor{mybluegray}{of a cyber incident} \end{tabular}};
    \node (risk) at (7.5,9) {\textbf{\Large{\textcolor{mybluegray}{Risk}}}};

\draw [myarrow] (mat_org) -- (2.5,1.2) -- (5,1.2) -- (prob_success);
\draw [myarrow] (compl_org) -- (7.5,1.2) -- (5,1.2) -- (prob_success);
\draw [myarrow2] (mat_agents) -- (prob_success);
\draw [myarrow] (prob_success) -- (likelihood);
\draw [myarrow] (num_threats) -- (likelihood);
\draw [myarrow2] (attr_org) -- (-2.5,4.5) -- (-2.5,5.8) -- (-1.5,5.8);
\draw [myarrow2] (attr_org) -- (-2.5,4.5) -- (-2.5,3) -- (-1.5,3);
\draw [myarrow] (-8.2,4.3) -- (-7.5,4.3) -- (-7.5,0) -- (mat_org);
\draw [myarrow3] (-8.2,4.7) -- (-7.5,4.7) -- (-7.5,6.2) -- (-1.5,6.2);
\draw [myarrow,mybluegray] (likelihood) -- (5,7.8) -- (7.5,7.8) -- (risk);
\draw [myarrow,mybluegray] (impact) -- (10,7.8) -- (7.5,7.8) -- (risk);

\end{tikzpicture}
}

%% file: Figures/varia_x0.tex
\resizebox{\columnwidth}{!}{
\begin{tikzpicture}
\begin{axis}[
xtick={0,1,2,...,10},
ytick={0, 0.1, 0.2,...,1},
xmin = 0,
xmax= 10,
grid = both,
ymin=0,
ymax=1,
legend style={at={(1.02,1)},anchor=north west},
mark size=2pt,
xlabel={$x$},
ylabel={$P_s$},
xticklabel style={
        /pgf/number format/fixed,
        /pgf/number format/precision=5
}
]

\addplot[blue, line width=0.7pt]coordinates{
(0,0.97)
(0.1,0.94414)
(0.2,0.91716)
(0.3,0.88915)
(0.4,0.86017)
(0.5,0.83034)
(0.6,0.79977)
(0.7,0.76859)
(0.8,0.73694)
(0.9,0.70498)
(1,0.67285)
(1.1,0.64073)
(1.2,0.60877)
(1.3,0.57712)
(1.4,0.54594)
(1.5,0.51537)
(1.6,0.48554)
(1.7,0.45656)
(1.8,0.42854)
(1.9,0.40157)
(2,0.37571)
(2.1,0.35101)
(2.2,0.32752)
(2.3,0.30526)
(2.4,0.28424)
(2.5,0.26444)
(2.6,0.24587)
(2.7,0.22849)
(2.8,0.21227)
(2.9,0.19716)
(3,0.18314)
(3.1,0.17014)
(3.2,0.15812)
(3.3,0.14703)
(3.4,0.1368)
(3.5,0.1274)
(3.6,0.11875)
(3.7,0.11083)
(3.8,0.10356)
(3.9,0.096912)
(4,0.090832)
(4.1,0.085278)
(4.2,0.080209)
(4.3,0.075587)
(4.4,0.071374)
(4.5,0.067538)
(4.6,0.064046)
(4.7,0.060869)
(4.8,0.057981)
(4.9,0.055356)
(5,0.052972)
(5.1,0.050807)
(5.2,0.048841)
(5.3,0.047057)
(5.4,0.045439)
(5.5,0.043971)
(5.6,0.04264)
(5.7,0.041433)
(5.8,0.040339)
(5.9,0.039347)
(6,0.038448)
(6.1,0.037634)
(6.2,0.036897)
(6.3,0.036229)
(6.4,0.035624)
(6.5,0.035076)
(6.6,0.034579)
(6.7,0.03413)
(6.8,0.033723)
(6.9,0.033355)
(7,0.033021)
(7.1,0.032719)
(7.2,0.032446)
(7.3,0.032199)
(7.4,0.031975)
(7.5,0.031772)
(7.6,0.031588)
(7.7,0.031422)
(7.8,0.031272)
(7.9,0.031136)
(8,0.031013)
(8.1,0.030902)
(8.2,0.030801)
(8.3,0.030709)
(8.4,0.030627)
(8.5,0.030552)
(8.6,0.030485)
(8.7,0.030423)
(8.8,0.030368)
(8.9,0.030318)
(9,0.030273)
(9.1,0.030232)
(9.2,0.030194)
(9.3,0.030161)
(9.4,0.03013)
(9.5,0.030103)
(9.6,0.030078)
(9.7,0.030056)
(9.8,0.030035)
(9.9,0.030017)
(10,0.03)
};\addlegendentry{$x_0=1$};

\addplot[red, line width=0.7pt]coordinates{
(0,0.97)
(0.1,0.96933)
(0.2,0.9686)
(0.3,0.96779)
(0.4,0.9669)
(0.5,0.96591)
(0.6,0.96482)
(0.7,0.96362)
(0.8,0.9623)
(0.9,0.96084)
(1,0.95924)
(1.1,0.95747)
(1.2,0.95553)
(1.3,0.95339)
(1.4,0.95104)
(1.5,0.94845)
(1.6,0.94561)
(1.7,0.94249)
(1.8,0.93906)
(1.9,0.93531)
(2,0.93119)
(2.1,0.92669)
(2.2,0.92176)
(2.3,0.91638)
(2.4,0.9105)
(2.5,0.9041)
(2.6,0.89713)
(2.7,0.88956)
(2.8,0.88134)
(2.9,0.87243)
(3,0.86281)
(3.1,0.85242)
(3.2,0.84123)
(3.3,0.82921)
(3.4,0.81633)
(3.5,0.80257)
(3.6,0.78791)
(3.7,0.77233)
(3.8,0.75584)
(3.9,0.73844)
(4,0.72014)
(4.1,0.70098)
(4.2,0.681)
(4.3,0.66024)
(4.4,0.63877)
(4.5,0.61667)
(4.6,0.59402)
(4.7,0.57093)
(4.8,0.54748)
(4.9,0.5238)
(5,0.5)
(5.1,0.4762)
(5.2,0.45252)
(5.3,0.42907)
(5.4,0.40598)
(5.5,0.38333)
(5.6,0.36123)
(5.7,0.33976)
(5.8,0.319)
(5.9,0.29902)
(6,0.27986)
(6.1,0.26156)
(6.2,0.24416)
(6.3,0.22767)
(6.4,0.21209)
(6.5,0.19743)
(6.6,0.18367)
(6.7,0.17079)
(6.8,0.15877)
(6.9,0.14758)
(7,0.13719)
(7.1,0.12757)
(7.2,0.11866)
(7.3,0.11044)
(7.4,0.10287)
(7.5,0.095897)
(7.6,0.089495)
(7.7,0.083621)
(7.8,0.078239)
(7.9,0.073313)
(8,0.068809)
(8.1,0.064694)
(8.2,0.060939)
(8.3,0.057514)
(8.4,0.054393)
(8.5,0.051551)
(8.6,0.048964)
(8.7,0.04661)
(8.8,0.044471)
(8.9,0.042526)
(9,0.04076)
(9.1,0.039156)
(9.2,0.037699)
(9.3,0.036378)
(9.4,0.035179)
(9.5,0.034091)
(9.6,0.033105)
(9.7,0.032211)
(9.8,0.0314)
(9.9,0.030666)
(10,0.03)
};\addlegendentry{$x_0=5$};

\addplot[black, line width=0.7]coordinates{
(0,0.97)
(0.1,0.96998)
(0.2,0.96996)
(0.3,0.96994)
(0.4,0.96992)
(0.5,0.9699)
(0.6,0.96987)
(0.7,0.96984)
(0.8,0.96981)
(0.9,0.96977)
(1,0.96973)
(1.1,0.96968)
(1.2,0.96963)
(1.3,0.96958)
(1.4,0.96952)
(1.5,0.96945)
(1.6,0.96937)
(1.7,0.96929)
(1.8,0.9692)
(1.9,0.9691)
(2,0.96899)
(2.1,0.96886)
(2.2,0.96873)
(2.3,0.96858)
(2.4,0.96841)
(2.5,0.96823)
(2.6,0.96803)
(2.7,0.9678)
(2.8,0.96755)
(2.9,0.96728)
(3,0.96698)
(3.1,0.96665)
(3.2,0.96628)
(3.3,0.96587)
(3.4,0.96542)
(3.5,0.96492)
(3.6,0.96438)
(3.7,0.96377)
(3.8,0.9631)
(3.9,0.96237)
(4,0.96155)
(4.1,0.96065)
(4.2,0.95966)
(4.3,0.95857)
(4.4,0.95736)
(4.5,0.95603)
(4.6,0.95456)
(4.7,0.95294)
(4.8,0.95116)
(4.9,0.94919)
(5,0.94703)
(5.1,0.94464)
(5.2,0.94202)
(5.3,0.93913)
(5.4,0.93595)
(5.5,0.93246)
(5.6,0.92863)
(5.7,0.92441)
(5.8,0.91979)
(5.9,0.91472)
(6,0.90917)
(6.1,0.90309)
(6.2,0.89644)
(6.3,0.88917)
(6.4,0.88125)
(6.5,0.8726)
(6.6,0.8632)
(6.7,0.85297)
(6.8,0.84188)
(6.9,0.82986)
(7,0.81686)
(7.1,0.80284)
(7.2,0.78773)
(7.3,0.77151)
(7.4,0.75413)
(7.5,0.73556)
(7.6,0.71576)
(7.7,0.69474)
(7.8,0.67248)
(7.9,0.64899)
(8,0.62429)
(8.1,0.59843)
(8.2,0.57146)
(8.3,0.54344)
(8.4,0.51446)
(8.5,0.48463)
(8.6,0.45406)
(8.7,0.42288)
(8.8,0.39123)
(8.9,0.35927)
(9,0.32715)
(9.1,0.29502)
(9.2,0.26306)
(9.3,0.23141)
(9.4,0.20023)
(9.5,0.16966)
(9.6,0.13983)
(9.7,0.11085)
(9.8,0.082835)
(9.9,0.055861)
(10,0.03)
};\addlegendentry{$x_0=9$};

\end{axis}
\end{tikzpicture}
}